\documentclass[acmsmall,screen,nonacm]{acmart}

\usepackage{booktabs}
\usepackage{multirow}
\usepackage{tabularx}
\usepackage{makecell}
\usepackage{lipsum}

\usepackage[most]{tcolorbox}
\usepackage{xcolor}
\definecolor{MyTeal}{HTML}{008573}
\definecolor{MyLightCyan}{HTML}{E0F8F3}

\usepackage{threeparttable}

\newtcolorbox{example}[1]{
    title={Example - #1}, 
    colback=MyLightCyan,
    colframe=MyTeal,
    colbacktitle=MyTeal,
    coltitle=white,
    fonttitle=\bfseries\small,
    boxrule=0.8pt,
    sharp corners,
    arc=3pt,
    left=4pt, right=4pt, top=4pt, bottom=4pt,
    before skip=10pt, after skip=10pt,
    breakable 
}

\newcolumntype{C}{>{\centering\arraybackslash}X}

\DeclareMathAlphabet{\mathcal}{OMS}{cmsy}{m}{n}

\AtBeginDocument{%
  }

\setcopyright{acmlicensed}
\copyrightyear{2026}
\acmYear{2026}
\acmDOI{XXXXXXX.XXXXXXX}
\acmConference[ISSTA '26]{The 35th ACM SIGSOFT International Symposium on Software Testing and Analysis}{October 3--9, 2026}{Oakland, CA, USA}
\acmISBN{978-1-4503-XXXX-X/2026/01}

\begin{document}

\title{AmbiBench: Benchmarking Mobile GUI Agents Beyond One-Shot Instructions in the Wild}

\author{Jiazheng Sun}
\email{jzsun24@m.fudan.edu.cn}

\author{Mingxuan Li}
\email{mingxuanli25@m.fudan.edu.cn}

\affiliation{%
  \institution{Fudan University}
  \city{Shanghai}
  \country{China}
}

\author{Yingying Zhang}
\email{zhangyy5522@mails.jlu.edu.cn}
\affiliation{%
  \institution{Jilin University}
  \city{Changchun}
  \country{China}
}

\author{Jiayang Niu}
\email{niujy24@m.fudan.edu.cn}

\author{Yachen Wu}
\email{yachenwu24@m.fudan.edu.cn}

\author{Ruihan Jin}
\email{jinrh24@m.fudan.edu.cn}

\author{Shuyu Lei}
\email{sylei25@m.fudan.edu.cn}

\author{Pengrongrui Tan}
\email{prrtan24@m.fudan.edu.cn}

\author{Zongyu Zhang}
\email{mspandif@gmail.com}

\author{Ruoyi Wang}
\email{ruoyiwang23@m.fudan.edu.cn}

\author{Jiachen Yang}
\email{jcyang24@m.fudan.edu.cn}

\author{Boyu Yang}
\email{byyang24@m.fudan.edu.cn}

\author{Jiacheng Liu}
\email{jiacliu25@m.fudan.edu.cn}

\author{Xin Peng}
\email{pengxin@fudan.edu.cn}

\affiliation{%
  \institution{Fudan University}
  \city{Shanghai}
  \country{China}
}

\authorsaddresses{%
  Authors’ Contact Information: 
  Jiazheng Sun and Mingxuan Li (Co-first authors), Fudan University, Shanghai, China; 
  Xin Peng (Corresponding author), Fudan University, Shanghai, China, \path{pengxin@fudan.edu.cn}.
}

\renewcommand{\shortauthors}{Sun et al.}

\begin{abstract}
Benchmarks are paramount for gauging progress in the domain of Mobile GUI Agents. In practical scenarios, users frequently fail to articulate precise directives containing full task details at the onset, and their expressions are typically ambiguous. Consequently, agents are required to converge on the user's true intent via active clarification and interaction during execution. However, existing benchmarks predominantly operate under the idealized assumption that user-issued instructions are complete and unequivocal. This paradigm focuses exclusively on assessing single-turn execution while overlooking the alignment capability of the agent. To address this limitation, we introduce AmbiBench, the first benchmark incorporating a taxonomy of instruction clarity to shift evaluation from unidirectional instruction following to bidirectional intent alignment. Grounded in Cognitive Gap theory, we propose a taxonomy of four clarity levels: Detailed, Standard, Incomplete, and Ambiguous. We construct a rigorous dataset of 240 ecologically valid tasks across 25 applications, subject to strict review protocols. Furthermore, targeting evaluation in dynamic environments, we develop MUSE (Mobile User Satisfaction Evaluator), an automated framework utilizing an MLLM-as-a-judge multi-agent architecture. MUSE performs fine-grained auditing across three dimensions: Outcome Effectiveness, Execution Quality, and Interaction Quality. Empirical results on AmbiBench reveal the performance boundaries of SoTA agents across different clarity levels, quantify the gains derived from active interaction, and validate the strong correlation between MUSE and human judgment. This work redefines evaluation standards, laying the foundation for next-generation agents capable of truly understanding user intent.
\end{abstract}

\ccsdesc[500]{Software and its engineering~Software testing and debugging}
\ccsdesc[500]{Computing methodologies~Mobile agents}
\ccsdesc[500]{General and reference~Evaluation}

\keywords{Interactive Mobile Agent Benchmark, Agent-based Evaluation}

\received{20 February 2007}
\received[revised]{12 March 2009}
\received[accepted]{5 June 2009}

\maketitle 

\section{Introduction}
\begin{figure}[!h]
  \centering
  \includegraphics[width=\linewidth]{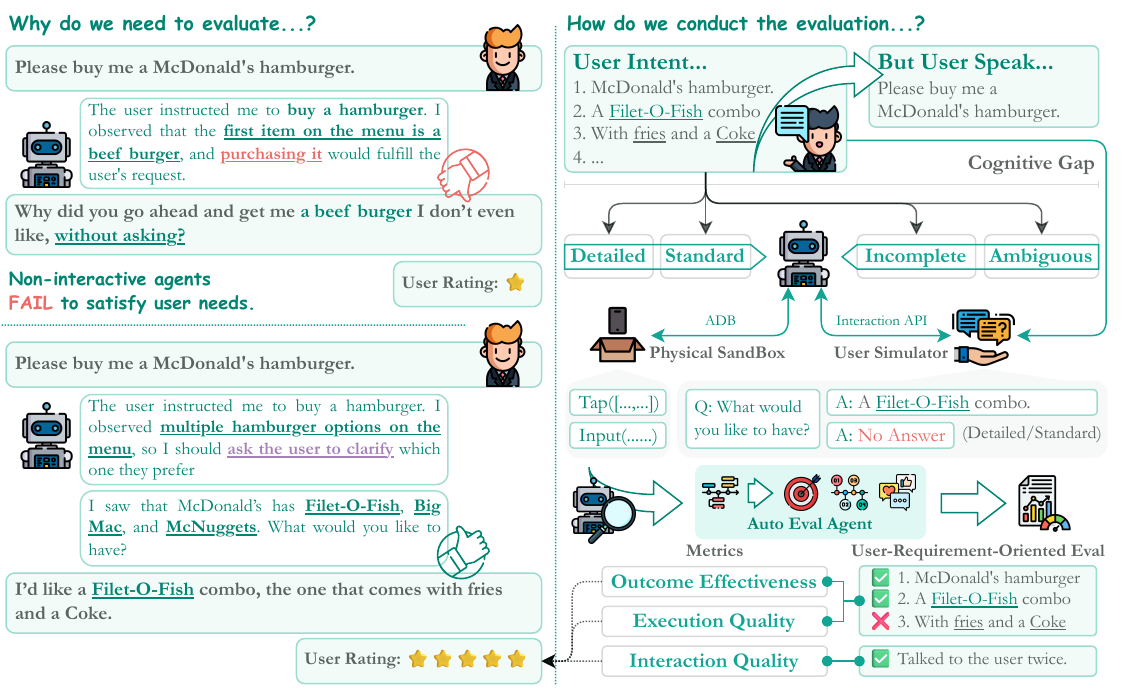}
  \caption{\textbf{Overview of the motivation and architecture of AmbiBench.} Non-interactive agents suffer from intent deviation due to probabilistic guessing (Top-Left), whereas interactive agents achieve intent alignment through active inquiry (Bottom-Left). To benchmark interactive agents, AmbiBench categorizes instructions into four clarity levels. A User Simulator, retaining Ground Truth information, engages in dynamic interaction with the agent, while the Sandbox synchronously captures execution traces. Subsequently, a multi-agent evaluation framework performs automated analysis, profiling agent capabilities at a fine granularity across three dimensions: Outcome Effectiveness, Execution Quality, and Interaction Quality.}
  \Description{}
  \label{figure 1}
\end{figure}

In recent years, driven by the continuous advancement of Large Language Models (LLMs) and Multimodal Large Language Models (MLLMs), research on Mobile GUI Agents has witnessed significant breakthroughs \cite{appagentx,appagentv2,cocoagent,Zhu_2025,nong2024mobileflow,ye2025mobileagentv3fundamentalagentsgui,cogAgent}. Ranging from the Agentic Workflow-based AppAgent\cite{appagent} to the Agent-as-a-Model UI-TARS \cite{qin2025uitarspioneeringautomatedgui}, these agents empower users to delegate complex operations via natural language instructions, exhibiting robust capabilities in task planning, screen perception, and device execution on mobile platforms. Consequently, reliable benchmarks for evaluating these agents are paramount for gauging progress in this domain.

However, existing research and benchmarks predominantly operate under the idealized assumption that user-issued instructions are complete and unequivocal—a premise we term Instruction equals Intent. Under this paradigm, evaluation focuses exclusively on quantifying execution accuracy, specifically assessing how faithfully an agent maps natural language instructions to sequences of UI actions. Yet, this One-Shot Instruction model overlooks the phenomenon of cognitive misalignment prevalent in real-world human-computer interaction. In practical scenarios, users frequently fail to articulate precise, system-level directives at the onset of a task; instead, their expressions are typically ambiguous or incomplete. For instance, a user might issue a vague command such as order a burger meal without specifying the burger type, sides, or beverages. Confronted with such instructions, existing non-interactive agents are often relegated to probabilistic guessing, inevitably failing to satisfy user requirements. Consequently, traditional evaluation frameworks driven solely by instruction adherence fail to assess the agent's alignment capability. specifically, its aptitude for leveraging active Interaction to converge on the user's true intent amidst information deficits.

While pioneering efforts have attempted to incorporate interaction mechanisms, the evaluation of interactive agents remains constrained by significant limitations. First, reliance on scripted responses severely restricts the generalization capability needed to handle unanticipated inquiries. Second, even state-of-the-art works incorporating dynamic simulators  employ coarse-grained evaluation strategies. They fail to systematically disentangle the differential impact of distinct clarity levels on agent performance and lack a fine-grained audit of interaction quality, such as information gain and inquiry validity. Consequently, existing evaluations struggle to discern whether an agent possesses genuine alignment capability or merely completes tasks through probabilistic guesswork. Furthermore, the absence of an explicit task clarity taxonomy in current benchmarks blurs the boundaries between execution, planning, and interaction capabilities, thereby impeding deep diagnosis and accurate error attribution regarding agent deficiencies.

To address these limitations, we present \textbf{AmbiBench, the first mobile benchmark to incorporate an instruction Clarity Level taxonomy}, thereby better reflecting the application usage patterns of real-world users. AmbiBench aims to precisely evaluate the agent's planning, execution, and interaction capabilities, promoting its transformation from one-way instruction following to two-way intent alignment. Figure \ref{figure 1} illustrates the motivation and architecture of it.

\textbf{Our first contribution} is the proposal of a task clarity taxonomy grounded in Cognitive Gap theory, upon which we construct an interactive evaluation framework. We systematically categorize instructions into four clarity levels (Detailed, Standard, Incomplete, and Ambiguous) and introduce an LLM-driven User Simulator that retains critical information to engage in dynamic interaction with the Mobile GUI Agent. This framework is designed to quantify the impact of varying task clarity on agent performance and to assess the agent's alignment capability via interaction. To the best of our knowledge, this represents the first dataset to categorize the clarity of task descriptions.

\textbf{Our second contribution} is the construction of AmbiBench, a diverse dataset comprising 240 tasks across 25 mainstream applications, accompanied by high-quality human trajectory annotations. Spanning both single-application and cross-application scenarios, the dataset is distinguished by its extensive inclusion of third-party applications. Distinct from prior benchmarks, AmbiBench pioneers strict Legitimacy review and Effectiveness assurance mechanisms. By eliminating meaningless posterior value-laden tasks and mitigating environmental noise, we ensure the Ecological Validity and robustness of the evaluation.

\textbf{Our third contribution} is the development of MUSE (Mobile User Satisfaction Evaluator), an automated evaluation framework enabling the MLLM-as-a-judge paradigm. By utilizing a standardized Interaction interface, MUSE achieves seamless integration with agents of diverse architectures—ranging from Agentic Workflow to Agent-as-a-Model. Leveraging a multi-agent adjudication architecture, it computes a suite of advanced, fine-grained metrics spanning the three dimensions of Outcome Effectiveness, Execution Quality, and Interaction Quality to precisely profile the comprehensive capabilities of the agent.

\textbf{Our fourth contribution} is a systematic empirical study designed to delineate the performance frontier of contemporary mobile agents. We conduct extensive evaluations across a spectrum of SoTA foundation models—ranging from general-purpose models (e.g., GPT-4o) to specialized fine-tuned models (e.g., UI-TARS\cite{qin2025uitarspioneeringautomatedgui}, AutoGLM\cite{liu2024autoglmautonomousfoundationagents}) as well as representative Agentic Workflows. Through this study, we deeply analyze the impact of instruction clarity levels on performance, quantify the gains yielded by active Interaction, and validate the high consistency between our automated metrics and human judgment.

In summary, this paper aims to redefine the evaluation standards for Mobile GUI Agents, \textbf{underscoring the criticality of achieving intent alignment through multi-turn interaction within uncertain environments}, thereby providing a solid foundation for constructing the next generation of agents that genuinely comprehend user intent.

\section{Related Work}
\subsection{Mobile GUI Agent}
A Mobile GUI Agent leverages LLMs as its primary reasoning engine to perceive and manipulate smartphone interfaces. Since its inception with AutoDroid \cite{AutoDroid}, the field has converged into two predominant technical paradigms: \textbf{1) Agent-as-a-Model} , characterized by specialized fine-tuning or architectural customization of (M)LLMs for device-level control (e.g., UI-TARS \cite{qin2025uitarspioneeringautomatedgui}, AutoGLM \cite{liu2024autoglmautonomousfoundationagents}); \textbf{2) Agentic Workflow}, which employs off-the-shelf foundation models within modular, pipeline-based frameworks (e.g., AppAgent \cite{appagent}, Mobile-Agent \cite{wang2024mobileagent,mobile-agent-v2,wang2025mobileagente}). Recent hybrid approaches, such as MobiAgent \cite{mobiagent}, seek to integrate the strengths of both paradigms.

While these agents exhibit proficiency in following one-shot instructions, they predominantly operate under the idealized assumption that the initial user intent is exhaustive and crystalline. This paradigm overlooks the reality of human-computer interaction, where requirements are often expressed progressively during execution. To transcend this limitation, pioneering efforts like Fairy \cite{sun2025robustobservableevolvableagentic} and the Uncertainty-Aware GUI Agent \cite{hao2025uncertaintyawareguiagentadaptive} have integrated user clarification mechanisms to mitigate the inherent ambiguity of initial directives. Nevertheless, from an evaluative standpoint, these advancements are currently hamstrung by data silos. The absence of standardized, publicly available interactive benchmarks compels researchers to rely on proprietary or small-scale datasets, thereby precluding equitable cross-comparisons and obstructing fine-grained error attribution within multi-turn Interaction trajectories.

\begin{table}[htbp]
\centering
\caption{Comprehensive Comparison of Mobile Benchmarks.}
\label{tab:benchmarks}
\scriptsize
\setlength{\tabcolsep}{1.2pt} 
\renewcommand{\arraystretch}{0.9}

\begin{threeparttable}

\begin{tabularx}{\textwidth}{
    l 
    >{\hsize=0.6\hsize\centering\arraybackslash}X 
    >{\hsize=0.6\hsize\centering\arraybackslash}X 
    >{\hsize=0.6\hsize\centering\arraybackslash}X 
    >{\hsize=0.5\hsize\centering\arraybackslash}X 
    >{\hsize=0.6\hsize\centering\arraybackslash}X 
    >{\hsize=1.6\hsize\centering\arraybackslash}X 
    >{\hsize=0.8\hsize\centering\arraybackslash}X 
    >{\hsize=1.3\hsize\centering\arraybackslash}X 
}
\toprule

\multirow{3}{*}[-1.5ex]{\textbf{Benchmark}} & 
\multirow{3}{*}[-1.3ex]{\makecell{\textbf{Comp.}\\\textbf{Support}}} & \multicolumn{4}{c}{\textbf{Scale}} & 
\multirow{3}{*}[-1.3ex]{\makecell{\textbf{Construction}\\\textbf{Pipeline}}} &
\multicolumn{2}{c}{\textbf{Task Settings}} \\

\cmidrule(lr){3-6} \cmidrule(lr){8-9}
& & \multirow{2}{*}[-0.5ex]{\textbf{\#Apps}} & 
\multicolumn{2}{c}{\textbf{\#Tasks}} & 
\multirow{2}{*}[-0.5ex]{\textbf{\#Steps}} & & 
\multirow{2}{*}[-0.5ex]{\makecell{\textbf{User}\\\textbf{Simulation}}} & 
\multirow{2}{*}[-0.5ex]{\makecell{\textbf{Instruction}\\\textbf{Taxonomy}}} \\
\cmidrule(lr){4-5}
& & & \textbf{Gen.} & \textbf{Inter.} & & & & \\
\midrule
AndroidArena \cite{xing2024understandingweaknesslargelanguage} & Microsoft & 13 & 221 & $\times$ & 7 & Auto-Gen (Function) & $\times$ & Single-level \\
AndroidWorld \cite{rawles2025androidworld} & Google & 20 & 116 & $\times$ & 8 & Auto-Gen (Template) & $\times$ & Single-level \\
MobileAgentBench \cite{wang2025mobileagentbench} & - & 10 & 100 & $\times$ & 4 & Human-Annotated & $\times$ & Single-level \\
MobileBenchV2 \cite{xu2025mobilebenchv2realisticcomprehensivebenchmark} & Xiaomi & 49 & \textbf{800*} & 100 & $\sim$7 & Auto-Gen (Trace) & ※ Presets & Binary (Subset) \\
SPABench \cite{chen2025spabench} & Huawei & \textbf{68} & 340 & $\times$ & 8 & Human-Annotated & $\times$ & Single-level \\
ColorBench \cite{song2025colorbenchbenchmarkingmobileagents} & OPPO & 21 & 175 & $\times$ & \textbf{13} & Human-Annotated  & $\times$ & Single-level \\
MobileWorld \cite{kong2025mobileworldbenchmarkingautonomousmobile} & Alibaba & 20 & 116 & 45 & n/a & Human-Annotated & \checkmark \textbf{Agent.} & Binary (Subset) \\
\midrule
\textbf{AmbiBench (Ours)} & - & 25 & 132 & \textbf{108} & 9 & Human-Annotated & \checkmark \textbf{Agent.} & \textbf{4 Levels} \\
\midrule[\heavyrulewidth] 
\end{tabularx}

\begin{tabularx}{\textwidth}{
    l 
    >{\hsize=0.5\hsize\centering\arraybackslash}X 
    >{\hsize=0.5\hsize\centering\arraybackslash}X 
    >{\hsize=0.5\hsize\centering\arraybackslash}X 
    >{\hsize=1.3\hsize\centering\arraybackslash}X 
    >{\hsize=1.2\hsize\centering\arraybackslash}X 
    >{\hsize=1.5\hsize\centering\arraybackslash}X 
    >{\hsize=1.5\hsize\centering\arraybackslash}X
}

\multirow{2}{*}[-0.8ex]{\textbf{Benchmark}} &
\multicolumn{3}{c}{\textbf{\#Metrics}} & 
\multirow{2}{*}[-0.8ex]{\textbf{Granularity}} & 
\multirow{2}{*}[-0.5ex]{\makecell{\textbf{Eval.}\\\textbf{Methodology}}} & 
\multirow{2}{*}[-0.5ex]{\makecell{\textbf{Auto Eval.}\\\textbf{Integration}}} & 
\multirow{2}{*}[-0.8ex]{\textbf{Environment}} \\
\cmidrule(lr){2-4}
& \textbf{Out.} & \textbf{Proc.} & \textbf{Inter.} & & & & \\
\midrule
AndroidArena \cite{xing2024understandingweaknesslargelanguage}& 1 & 3 & $\times$ & Bin / Step & Traj. Match & $\times$ Ad-hoc & Local App\\
AndroidWorld \cite{rawles2025androidworld}& 1 & $\times$ & $\times$ & Binary & State Check & \checkmark Sandbox (Orch.) & Local App\\
MobileAgentBench \cite{wang2025mobileagentbench}& 1 & 2** & $\times$ & Binary & State Check & \checkmark Sandbox (Orch.) & Local App\\
MobileBenchV2 \cite{xu2025mobilebenchv2realisticcomprehensivebenchmark} & 1 & 3 & $\times$ & Bin / Step & Traj. Match & $\times$ Ad-hoc & Offline Dataset\\
SPABench \cite{chen2025spabench}& 1 & 2** & $\times$ & Bin / Step & VLM Adjud. & \checkmark \textbf{Sandbox (Proxy)} & Online App\\
ColorBench \cite{song2025colorbenchbenchmarkingmobileagents}& \textbf{2} & 1 & $\times$ & Bin / Task & Traj. Match & ※ VLM-Friendly Frm. & Offline Dataset\\
MobileWorld \cite{kong2025mobileworldbenchmarkingautonomousmobile}& 1 & 1 & \textbf{2} & Bin / Step & State Check & ※ VLM-Friendly Frm. & Local App\\
\midrule
\textbf{AmbiBench (Ours)} & \textbf{2} & \textbf{3**} & \textbf{2} & \textbf{Req/Bin/Step} & VLM Adjud.& \checkmark \textbf{Sandbox (Proxy)} & Online App\\
\bottomrule
\end{tabularx}

\begin{tablenotes}
    \footnotesize
    \item[*] Only the 800 tasks actually used for testing (out of the total 12,856) are counted.
    \item[**] Homogeneous metrics are consolidated to align with other works (e.g., multiple metrics related to abnormal termination are counted as a single indicator).
\end{tablenotes}

\end{threeparttable}
\end{table}

\subsection{Mobile GUI Agent Benchmarks}

As shown in Table \ref{tab:benchmarks}, existing evaluation methodologies are broadly categorized into \textbf{Dataset-based Offline Static Evaluation} \cite{xu2025mobilebenchv2realisticcomprehensivebenchmark,MobileM3,li2025autoguiscalingguigrounding}  and \textbf{Online Dynamic Evaluation} \cite{mobile—env,androidlab,rawles2025androidworld,xing2024understandingweaknesslargelanguage,wang2025mobileagentbench,chen2025spabench,MVISU-Bench,a3}. Early static evaluations, exemplified by Mobile3M \cite{MobileM3} and MobileBenchV2 \cite{xu2025mobilebenchv2realisticcomprehensivebenchmark}, were predicated on Trajectory Chains but suffered from the inherent limitation of Path Rigidity. To circumvent this, recent studies such as ColorBench \cite{song2025colorbenchbenchmarkingmobileagents} have adopted Trajectory Graphs to accommodate the diversity of agent decision-making. In pursuit of higher Ecological Validity, dynamic benchmarks have bifurcated into two streams: local utility applications (e.g., AndroidArena \cite{xing2024understandingweaknesslargelanguage}, AndroidWorld \cite{rawles2025androidworld}, MobileAgentBench \cite{wang2025mobileagentbench}) and web-based dynamic environments (e.g., SPA-Bench \cite{chen2025spabench}, Mobile-Eval-E \cite{wang2025mobileagente}). While the latter approximate real-world scenarios more closely, they face severe challenges regarding state determinism: stochastic content, such as live feeds and advertisements—renders Ground, truth Verification non-trivial, a process otherwise feasible within controllable local applications. Currently, the field confronts three converging threats: task saturation obscures true performance disparities among SoTA models; prohibitive data scales combined with ad-hoc evaluation suites raise barriers to reproducibility; and suboptimal task design coupled with coarse-grained binary accuracy metrics severely limits analytical depth. To mitigate these deficiencies, AmbiBench introduces a rigorous cognitive filtering mechanism to ensure data quality, alongside a standardized automated evaluation suite and multi-dimensional, fine-grained metrics.

\subsection{User Interaction Benchmarks}
While multi-turn interaction is pivotal for resolving real-world ambiguities, most existing benchmarks remain confined to static evaluations. For instance, the interaction subset in MobileBenchV2 \cite{xu2025mobilebenchv2realisticcomprehensivebenchmark} relies on rigid, scripted responses tied to predefined trajectory nodes, thereby lacking the generalization necessary for dynamic scenarios. Although RealEval \cite{sun2025robustobservableevolvableagentic} and the concurrent work MobileWorld \cite{kong2025mobileworldbenchmarkingautonomousmobile} have pioneered the integration of LLM-based agentic user simulators to enable dynamic responsiveness, their evaluative strategies remain relatively coarse-grained. Specifically, contemporary frameworks suffer from two primary limitations: first, they fail to systematically decouple the impact of varying clarity levels on agent performance, which precludes precise error attribution; second, they lack a rigorous audit of interaction quality, making it nearly impossible to discern whether an agent’s success stems from meaningful communication or stochastic guessing. To bridge this evaluative chasm, AmbiBench introduces a novel four-level instruction clarity taxonomy grounded in the Gulf of Envisioning and establishes a multi-dimensional interaction auditing suite. Leveraging an MLLM-as-a-judge paradigm, our framework performs a holistic semantic audit of both execution efficacy and interaction quality from a human-centric cognitive perspective.

\section{AmbiBench}
\subsection{Task Formulation}

\subsubsection{Task Modeling}

We conceptualize the mobile task as a process of information transmission from the user cognitive space to the system parameter space. Throughout this transmission, a reduction in information entropy is inevitable. Accordingly, we model the task as a decoupled tuple comprising the \textbf{Ground Truth Intent} $\mathcal{U}_{gt}$ and the \textbf{Observed Instruction} $\mathcal{I}_{obs}$.
\begin{enumerate}
    \item \textbf{Ground Truth Intent $\mathcal{U}_{gt}$:} The set of user requirements within a specific context. It consists of a series of determinate, indivisible atomic requirements: $\mathcal{U}_{gt} = \{r_1, r_2, ..., r_n\}$. 
Here, each $r_i$ denotes a specific constraint or expectation regarding the task outcome. To enable a more granular analysis of the necessity for Interaction, we further classify atomic requirements into three distinct categories:
 \textbf{a) Anchor Requirements $r_{anchor}$:} The core objects that determine the fundamental type of the task.
 \textbf{b) Explicit Constraints $r_{explicit}$:} Parameters explicitly displayed by the system that require user selection.
 \textbf{c) Implicit Preferences $r_{implicit}$:} Parameters that are collapsed or nested within the system interface and possess default values, yet remain modifiable for user personalization.
Taking a McDonald's ordering task as an illustrative example, the user's Ground Truth Intent $\mathcal{U}_{gt}$ may comprise the following three atomic requirements: a) $r_{anchor}$ : The user intends to order a Filet-O-Fish Meal. b) $r_{explicit}$ : The user specifies a medium Coke Zero as the beverage. c) $r_{implicit}$ : The user prefers the Coke to be served without ice.
\item \textbf{Observed Instruction $\mathcal{I}_{obs}$:} This represents the natural language description received by the agent. Due to linguistic constraints or user expressive limitations, $\mathcal{I}_{obs}$ typically manifests as an incomplete subset of the Ground Truth Intent  $\mathcal{U}_{gt}$. We formally define the Cognitive Gap $\mathcal{G}$ as the set of requirements not explicitly encompassed by the instruction: $\mathcal{G} = \mathcal{U}_{gt} \setminus \mathcal{S}(\mathcal{I}_{obs})$, where $\mathcal{S}(\mathcal{I}_{obs})$ denotes the atomic requirements explicitly extracted from the instruction. 
\end{enumerate}
AmbiBench advocates a requirement-satisfaction-oriented paradigm to evaluate agent proficiency, where the evaluative focus dynamically shifts based on the presence of a Cognitive Gap. In scenarios where $\mathcal{G} = \emptyset$, the Requirement Coverage Rate (RCR) serves as a proxy for execution performance, assessing whether the agent can faithfully fulfill all explicit directives. Conversely, when $\mathcal{G} \neq \emptyset$, signaling the presence of uncommunicated implicit requirements $r \in \mathcal{G}$, the RCR further quantifies the agent's ability to retrieve vital incremental information $\Delta I$ via interactive actions $a_{ask}$. Irrespective of initial clarity levels, the terminal objective of the AmbiBench evaluation remains steadfastly anchored in the ultimate fulfillment of user intent, transcending mere superficial instruction-following.

\subsubsection{Taxonomy: Clarity Hierarchy}
To precisely decouple the agent's \textbf{Execution}, \textbf{Planning}, and \textbf{Interaction} capabilities, AmbiBench establishes a four-level clarity taxonomy grounded in the Gulf of Envisioning theory \cite{chi24}. This taxonomy categorizes scenarios based on the specificity with which the instruction $\mathcal{I}_{obs}$ covers the Ground Truth Intent $\mathcal{U}_{gt}$, as well as the presence of core anchors.
\begin{enumerate}
    \item \textbf{Detailed (Over-defined):} $\mathcal{I}_{obs}$ constitutes a Procedural Instruction that not only fully encompasses $\mathcal{U}_{gt}$ but also explicitly encodes the system-level UI operation path $\tau_{ref}$.
    At this point, $\mathcal{S}(\mathcal{I}_{obs}) \supseteq \mathcal{U}_{gt} \cup \tau_{ref}$, and $\mathcal{G} = \emptyset$. 
    This level substantially mitigates the necessity for long-horizon Planning and memory retrieval, effectively reducing the agent's role to that of a mere instruction executor. Its primary objective is to assess the agent's proficiency in instruction adherence and execution.
    \begin{example}{Detailed}
    Open the McDonald's App, navigate to the menu page, and search for Filet-O-Fish Meal. Within the beverage options, select Medium Coke Zero, then choose No Ice under the Customize section. Add the item to the cart, proceed to checkout, and pay for the order.
    \end{example}

    \item  \textbf{Standard (Well-defined):} $\mathcal{I}_{obs}$ constitutes a Goal-Oriented Instruction that fully encompasses $\mathcal{U}_{gt}$ yet omits explicit operational steps. This represents the ideal paradigm for agent interaction. At this point, $\mathcal{S}(\mathcal{I}_{obs}) \supseteq \mathcal{U}_{gt}$, and $\mathcal{G} = \emptyset$. At this level, the agent is required to autonomously derive the optimal trajectory from the initial state $s_0$ to a state satisfying $\mathcal{U}_{gt}$, primarily targeting the evaluation of the agent's Planning capabilities.
    \begin{example}{Standard}
    Search for Filet-O-Fish Meal in the McDonald's app, select Medium Zero-Calorie Coke, choose No Ice, add it to your cart, and pay for the order.
    \end{example}
    \item \textbf{Incomplete (Under-defined):} At this level, the observed instruction exhibits an Instruction Gap, characterized by the presence of core anchor requirements but the omission of specific explicit or implicit constraints. Formally, this is denoted as $r_{anchor} \in \mathcal{S}(\mathcal{I}_{obs})$ and $\mathcal{G} \cap \{r_{explicit}, r_{implicit}\} \neq \emptyset$. Within this paradigm, the agent is compelled to dynamically identify missing parameters (slots) during Execution and initiate Interaction for slot-filling. This level is specifically designed to stress-test the agent's proficiency in proactive disambiguation and its Alignment Capability.
    \begin{example}{Incomplete}
    Order a Filet-O-Fish Meal in the McDonald's app.
    \end{example}
    \item \textbf{Ambiguous (Ill-defined):} At this level, the observed instruction suffers from an Intentionality Gap, characterized by a vague high-level goal that lacks the core anchor requirement. Formally, this is expressed as $r_{anchor} \in \mathcal{G}$. Consequently, the task space remains unconverged, encompassing multiple divergent anchor candidates. Under these circumstances, the agent is precluded from immediate Planning; instead, it must facilitate intent clarification by presenting heuristic options or posing guided inquiries to help the user prune the search space. This scenario serves as a rigorous evaluation of the agent’s sophisticated Interaction and Alignment Capability in highly underspecified contexts.
    \begin{example}{Ambiguous}
    Order a hamburger meal.
    \end{example}
\end{enumerate}

\subsection{Dataset Construction}
\subsubsection{Task Legitimacy and Effectiveness Assurance} \label{section 3.2.1}
To safeguard Ecological Validity, AmbiBench rigorously delineates task boundaries by leveraging a two-dimensional taxonomy matrix characterized by Requirement Nature and Intent Origin.

From the perspective of \textbf{Requirement Nature}, we categorize tasks into: \textbf{1) Functional Requirements:} These entail operational objectives centered on efficiency and objective outcomes (e.g., \textit{food ordering, navigation, and system configuration}). Characterized by minimal social risk, such tasks are well-suited for delegation to autonomous agents. \textbf{2)Value-Laden Requirements:} These involve operations that implicate a user’s Social Reputation and Affective Stance (e.g., \textit{liking, commenting, reposting, or direct messaging}). Due to their high degree of subjectivity and potential social ramifications, these actions impose a stringent threshold for authorized delegation.

Along the dimension of \textbf{Intent Origin}, we distinguish between: \textbf{1) Prior Intent:} Requirements originating from a user’s prior experience or established memory. In this context, the user possesses distinct Epistemic Certainty regarding the target entity—i.e., they know the target’s existence prior to Interaction (e.g., \textit{Navigate to a specific store}). \textbf{2) Posterior Intent:} Requirements that are dynamically formulated and refined based on real-time information acquisition. Here, the user operates under a state of information deficit, where the primary objective is exploration-before-decision (e.g., \textit{See what shops are nearby}). Consequently, the intent progressively converges as the interaction unfolds.

\begin{figure}[h]
  \centering
  \includegraphics[width=\linewidth]{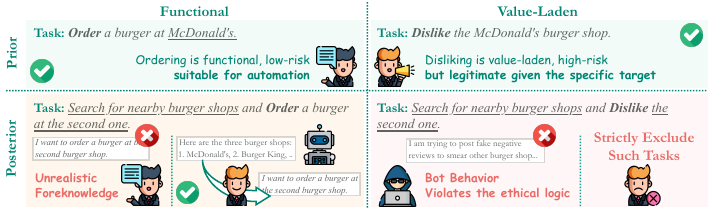}
  \caption{\textbf{Classification matrix for task legitimacy and effectiveness assurance.} This matrix establishes task admission and transformation logic across two dimensions: Requirement Nature (Functional/Value-Laden) and Intent Origin (Prior/Posterior). \textbf{Case A (Retain):} Tasks characterized by Prior Intent align with human cognitive logic and possess a determinate set of requirements; thus, AmbiBench retains these in their entirety. \textbf{Case B (Convert):} For Posterior Intent tasks, users cannot foresee outcomes ab initio, rendering it impossible to articulate plausible Detailed or Standard instructions. However, they can naturally issue Incomplete or Ambiguous instructions. For such scenarios, AmbiBench employs a Preset Posterior transformation strategy, which populates user requirements with predefined selections to reconstruct the task as a determinate, pseudo-prior objective. \textbf{Case C (Exclude):} In the absence of actual content consumption, users would implicitly refuse to authorize an agent to articulate an Affective Stance on their behalf. Consequently, posterior value-laden tasks contravene the ethical logic governing intelligent agents, and AmbiBench rigorously excludes them.}
  \Description{}
  \label{figure 2}
\end{figure}

Predicated on the intersection of these dimensions, we have established differentiated protocols for Ground-Truth $\mathcal{U}_{gt}$ construction as shown in Figure \ref{figure 2}.

To mitigate systemic False Positives arising from agents capitalizing on system defaults and False Negatives induced by environmental unreliability, AmbiBench incorporates two core mechanisms designed to safeguard evaluative Effectiveness and robustness:
\begin{enumerate}
    \item \textbf{Forced Non-Default Principle:} Preset default configurations in UI controls often allow suboptimal agents to fulfill tasks through stochastic actions, thereby masking inherent deficiencies in their reasoning capabilities. To eliminate this speculative margin, we employ an Adversarial Strategy during $\mathcal{U}_{gt}$ construction: all parameters susceptible to omission in the instruction are mandated to non-default values (e.g., \textit{if the system defaults to Coke, the Ground-Truth is  set to Coke Zero}). This mechanism compels the agent to proactively initiate Interaction to resolve requirements, precluding reliance on serendipity or system defaults.
    \item \textbf{Ensuring Alignment:} Given that mobile task execution is contingent upon the internal state of the application, we mitigate environmental noise—invalid evaluations stemming from inadequate provisioning—by developing companion ADB State Injection Scripts for each task. Prior to task initiation, these scripts automatically populate requisite data and configure system settings based on $\mathcal{U}_{gt}$ dependencies. This protocol ensures the Executability and State Consistency of each evaluation instance at the logical level.
\end{enumerate}

\subsubsection{Construction Pipeline} \label{section 3.2.2}
Departing from the conventional instruction-collection and path-annotation paradigm, AmbiBench pioneers a Requirement-First, Reverse Derivation construction strategy. This methodology primarily anchors a determinate list of Ground-Truth $\mathcal{U}_{gt}$ within real-world contexts; subsequently, by systematically modulating the decay of information entropy, we synthetically derive Observed Instructions $\mathcal{I}_{obs}$ across varying clarity levels. This pipeline is structured into four distinct stages:

\textbf{Phase 1: Requirement Anchoring.} To construct a Ground-Truth $\mathcal{U}_{gt}$ candidate pool with high ecological validity, we analyzed authentic user directives from public datasets such as VoiceBench and AITW. This analysis yielded meta-requirements across five pivotal domains: E-commerce, Social Platforms, Productivity \& Collaboration, Device System Management, and Information Retrieval. Within these domains, we performed functional tree traversals across 25 mainstream applications to extract two categories of atomic requirements: \textbf{a) Core Functional Requirements:} These encompass standard requirement sets derived from high-frequency scenarios, such as \textit{Food Ordering in McDonald's or Instant Messaging in WeChat}. \textbf{b) Long-tail Functional Requirements:} These involve non-core, low-frequency demands buried within the nested menus of Super Apps. These unguided tasks are specifically curated to probe the depth of an agent's semantic understanding. Examples include Ordering Prepaid Gift Cards in McDonald's or Configuring Auto-Translation in WeChat.

\textbf{Phase 2: Complexity Injection.} Since raw atomic requirements often lack sufficient depth, we infuse temporal and environmental constraints at the Ground-Truth $\mathcal{U}_{gt}$ level to simulate real-world sophistication: \textbf{a) Cross-App Contextual Chaining:} By leveraging inter-application complementarity, we concatenate atomic requirements from disparate apps into a singular logical workflow. This forces the agent to navigate state-retention challenges inherent in context switching. A representative scenario involves bridging content consumption with social sharing, such as \textit{Copy a song link from NetEase Cloud Music and forward it to a specific WeChat contact}. \textbf{b) Parameter Specification and Non-Defaulting:} We explicitly inject constraint parameters and rigorously enforce the Forced Non-Default Principle. This ensures that the agent is compelled to validate requirements via Interaction rather than bypassing the reasoning process by relying on system defaults or luck.

\textbf{Phase 3: Requirement Homogenization and Heterogenization.} To probe the upper bounds of an agent's generalization and discrimination, we implement two adversarial data augmentation strategies on the $\mathcal{U}_{gt}$ set: \textbf{a) Heterogenization for Generalization:} We map a singular high-level intent (e.g., \textit{Purchasing Coffee}) onto multiple, structurally disparate application environments (e.g., \textit{Luckin Coffee vs. Starbucks}). This generates a cluster of requirement sets that are semantically equivalent yet structurally heterogeneous, thereby validating the agent's Zero-Shot Transfer capabilities across diverse UI layouts. \textbf{b) Homogenization for Discrimination:} Within a high-density application environment, we construct requirement pairs that are semantically proximal but operationally distinct (e.g., Follow Author vs. Like Article). These tasks often involve UI elements with high visual and spatial overlap, enabling a precise evaluation of the agent’s fine-grained alignment and discrimination accuracy under confounding conditions.

\begin{figure}[h]
    \centering

    \begin{minipage}[t]{0.48\linewidth}
        \centering
        \includegraphics[height=3.9cm, keepaspectratio]{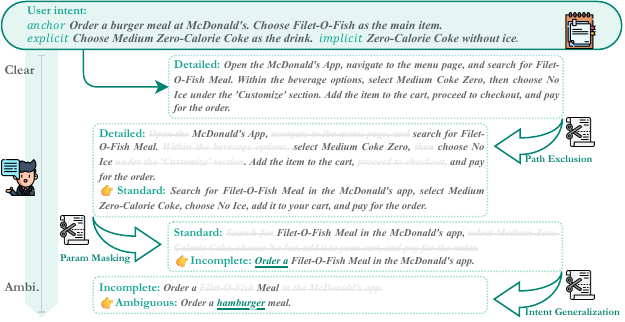}
        \caption{\textbf{The derivation process of task instructions across 4 clarity levels from requirements.}}
        \Description{}
        \label{figure 5}
    \end{minipage} \hfill
    \begin{minipage}[t]{0.48\linewidth}
        \centering
        \includegraphics[height=3.9cm, keepaspectratio]{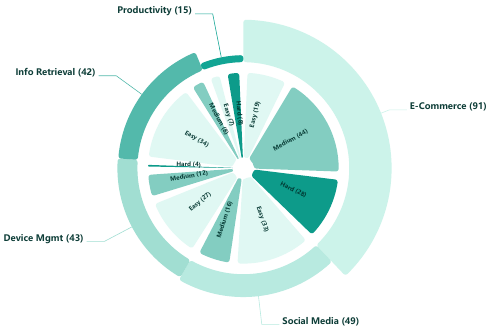}
        \caption{\textbf{Domain distribution and difficulty quantification statistics of the AmbiBench dataset.}}
        \Description{}
        \label{figure 6}
    \end{minipage}
\end{figure}

\textbf{Phase 4: Instruction Derivation via Cognitive Gaps.} Building upon the finalized $\mathcal{U}_{gt}$, we categorize user requirements into two functional subsets based on intent-dependency logic: the \textbf{Anchor Set} ($r_{anchor}$), which defines the existence of a task (whose absence renders the task type unidentifiable), and the \textbf{Parameter Set} ($r_{explicit/implicit}$), which defines the precision of a task (whose absence results in execution deviations). Finally, we employ a mechanism of controlled Information Stripping to systematically derive observed instructions across four distinct clarity levels, as shown in Figure \ref{figure 5}:

\begin{itemize}
    \item \textbf{Detailed (Full-Path Serialization):} In this stage, no information stripping is performed. We synthesize Procedural Instructions by integrating all anchors and parameters from $\mathcal{U}_{gt}$ with manually recorded Golden Traces $\tau_{ref}$. This results in a step-by-step directive that explicitly guides the agent through the UI.
    \item \textbf{Standard (Path Exclusion):} This stage implements De-proceduralization. While maintaining all anchors and parameter constraints from $\mathcal{U}_{gt}$, we systematically excise any descriptions pertaining to UI interaction steps. The resulting instruction specifies only the desired End State, compelling the agent to autonomously plan the optimal execution trajectory.
    \item \textbf{Incomplete (Parameter Masking):} This stage executes Parameter Masking. We preserve the $r_{anchor}$ but purposefully strip $r_{explicit}$ or $r_{implicit}$ from the instruction text. This necessitates Interaction for the agent to recover the missing requirements. Atomic tasks lacking parameter constraints bypass this level to avoid redundant samples.
    \item \textbf{Ambiguous (Intent Generalization):} This stage employs Hypernym Substitution. We replace specific $r_{anchor}$ terms with broader semantic categories and remove all restrictive modifiers. This engineering of a vast Intent Search Space forces the agent to facilitate multi-turn disambiguation.
\end{itemize}

\subsection{Dataset Statistics}
Figure \ref{figure 6} summarizes the key statistics of AmbiBench. The dataset comprises 240 evaluation tasks spanning 7 system applications and 18 third-party applications. These tasks cover five core domains of mobile usage: E-commerce (91), Social Platforms (49), Productivity \& Collaboration (15), Device System Management (43), and Information Retrieval (42). Notably, 108 tasks are classified as interactive tasks that necessitate user Interaction for completion. In total, AmbiBench defines 175 atomic user requirements. Stratified by difficulty based on the number of constituent atomic requirements, the dataset includes \textbf{120} Simple tasks (1–2 requirements), \textbf{80} Medium tasks (3–4 requirements), and \textbf{40} Hard tasks (5 or more requirements).

\section{MUSE (Mobile User Satisfaction Evaluator)}
While AmbiBench offers a testing environment with high Ecological Validity, its inherent environmental dynamism and interaction uncertainty pose significant challenges for automated evaluation. In Offline Static Evaluation (e.g., \textit{MobileBenchV2}), assessment merely requires matching the agent's predicted actions against pre-recorded Ground Truth trajectories. Similarly, in Online Dynamic Evaluation involving Local Apps within closed environments (e.g., \textit{AndroidWorld}), hard-coded deterministic State Checks remain viable due to the high controllability of the environment. However, in In-the-Wild evaluations represented by SPABench and AmbiBench, agents interact directly with Online Apps. The highly dynamic nature of web content renders traditional hard-coded rules ineffective, making the precise verification of requirement satisfaction difficult.

Furthermore, AmbiBench's expansion of the evaluation scope to include Interaction Quality introduces additional complexities. Determining the validity of inquiries, such as distinguishing between effective guidance and disruptive noise—relies on complex, context-dependent semantic reasoning rather than simple rule-based matching. To address these challenges, we propose MUSE (Mobile User Satisfaction Evaluator). Adopting the MLLM-as-a-judge paradigm\cite{llm-as-a-judge}, this framework employs a multi-agent architecture to simulate the cognitive perspective of human users, performing a semantic audit of the agent's execution effectiveness and interaction quality within dynamic environments.

\subsection{Multi-Agent Adjudication Architecture}
\subsubsection{Phase I: Dynamic Execution \& Data Acquisition }

\begin{figure}[h]
  \centering
  \includegraphics[width=0.82\linewidth]{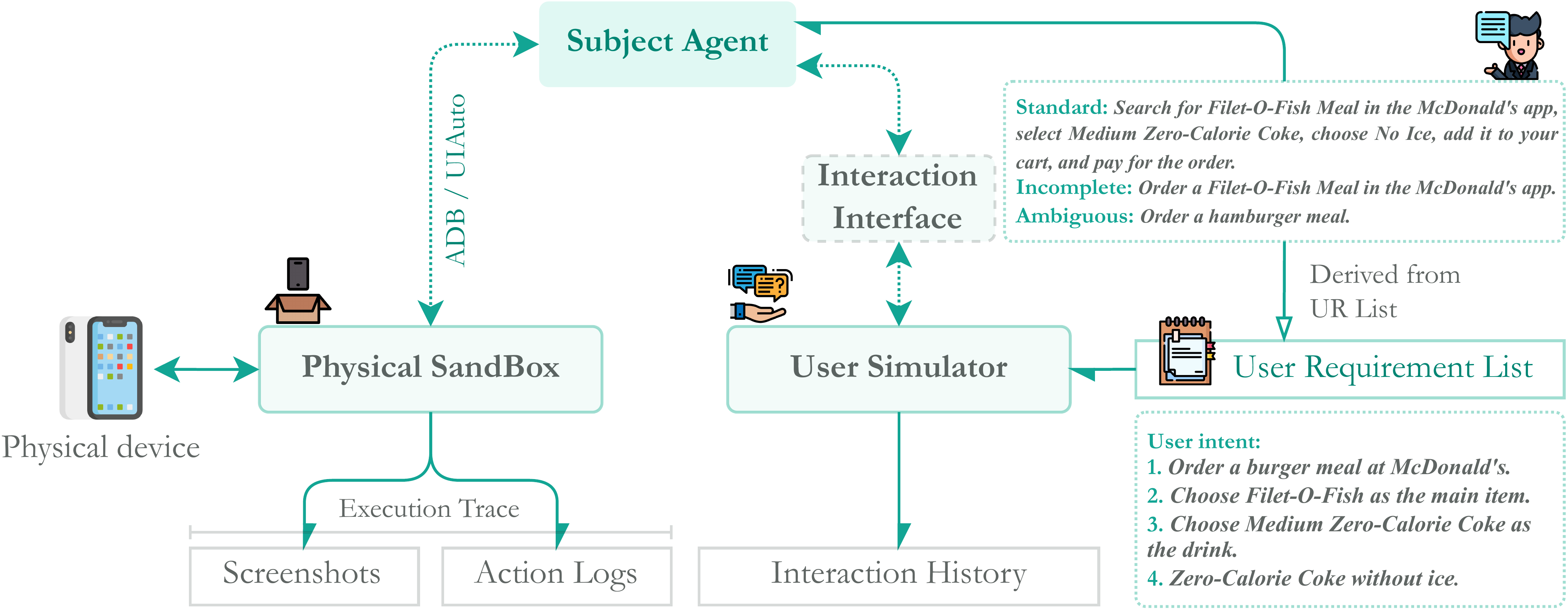}
  \caption{\textbf{MUSE data acquisition pipeline.} The system decouples the Subject Agent from the evaluation environment via a standardized interface: the Physical Sandbox executes instructions and captures execution traces containing screenshots and action logs, while the User Simulator answers the Subject Agent based on the requirement list and records the interaction history, collectively completing the collection of eval data.}
  \Description{}
\end{figure}
In this phase, MUSE dispatches the task instruction $\mathcal{I}_{obs}$ to the Subject Agent and comprehensively captures multimodal data throughout the execution process.
\begin{enumerate}
    \item \textbf{Interaction Interface:} To achieve architectural decoupling between the evaluation system and heterogeneous subject agents, MUSE defines a standardized Interaction Interface (HTTP API). Subject agents are required to conform to this interface to retrieve the initial instruction $\mathcal{I}_{obs}$. Furthermore, if an agent possesses conversational capabilities, it must utilize this channel to forward natural language queries and receive feedback from the User Simulator, thereby ensuring standardized interactive integration.
    \item \textbf{User Simulator:} Acting as a dynamic agent holding the Ground Truth Intent $\mathcal{U}_{gt}$, the User Simulator is tasked with responding to agent inquiries. Its behavior is rigorously governed by both $\mathcal{U}_{gt}$ and the task's clarity level:
    \textbf{a)} For informationally complete tasks (Detailed/Standard) or inquiries pertaining to low-level UI operations (e.g., \textit{Should I click?}), the simulator issues a standardized refusal: \textit{Please make your own decisions based on the current instructions.}
    \textbf{b)} Exclusively within Incomplete or Ambiguous tasks, the simulator queries $\mathcal{U}_{gt}$ to retrieve ground truth parameter values for its response; for parameters undefined in the intent, it returns \textit{No Preference.} The entire dialogue process is logged in a structured format as the Interaction History $H_{int}$.
    \item \textbf{Physical Sandbox:} MUSE constructs a proxy sandbox employing a Man-in-the-Middle (MitM) architecture. It exposes a standard ADB interface to abstract underlying device heterogeneity, enabling subject agents to integrate seamlessly without code modification. The sandbox backend bridges to a device farm composed of real physical devices or high-fidelity emulators. Prior to task initiation, the sandbox initializes App states based on task metadata (e.g., \textit{clearing shopping carts or logging into specific accounts}) to eliminate environmental noise. As the subject agent issues ADB commands, the system employs traffic sniffing to non-invasively capture the Execution Trace $\tau_{exec}$, which comprises sequential screenshots and atomic action logs.
\end{enumerate}
\textbf{Phase Output:} Upon conclusion, this phase yields a structured data packet encapsulating the full context, denoted as $\langle H_{int}, \tau_{exec} \rangle$. Serving as immutable raw evidence, this artifact is transferred to the subsequent phase for offline adjudication.

\subsubsection{Phase II: Automated Analysis \& Adjudication}
Upon the conclusion of the dynamic execution phase, MUSE initiates the offline analysis workflow. This process commences with the Trajectory Serialization Agent, which transforms the raw multimodal trajectory data into a semantic trajectory. This structured data is subsequently transferred to three decoupled Adjudication Agents for quantitative evaluation.
\begin{figure}[h]
  \centering
  \includegraphics[width=\linewidth]{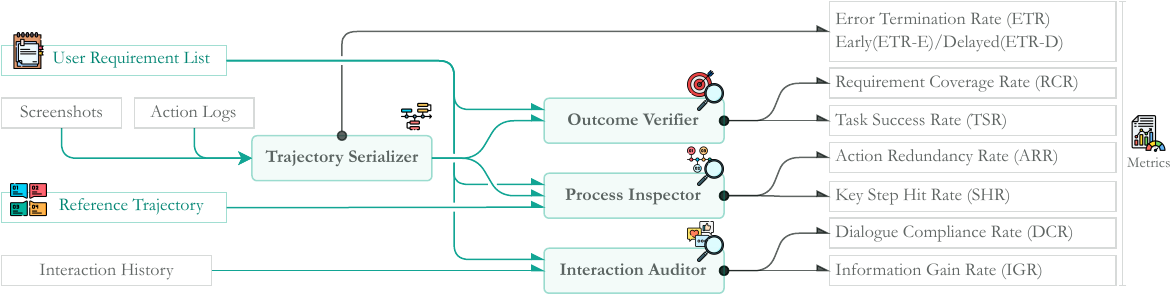}
  \caption{\textbf{MUSE quantitative evaluation pipeline based on multi-agent auditing.} The Trajectory Serializer first transforms raw multimodal execution traces into semantic structures. Subsequently, it coordinates with three specialized adjudication agents working in parallel: the Outcome Verifier validates task fulfillment based on the requirement list, the Process Inspector evaluates execution conformity by comparing against reference trajectories, and the Interaction Auditor quantifies information acquisition efficiency based on the interaction history, collectively generating comprehensive evaluation metrics.}
  \Description{}
\end{figure}

\textbf{Trajectory Serializer} $\mathcal{A}_{TS}$ derives the semantic trajectory sequence ($\tau_{exec}^\mathcal{S}$) from the raw execution trace ($\tau_{exec}$). This process integrates discrete operational and visual streams while simultaneously compressing high-dimensional multimodal data and mitigating non-semantic noise, ultimately yielding a semantically coherent, structured trajectory:$$\tau_{exec}^\mathcal{S} = \langle e_1, e_2, ..., e_t \rangle = \mathcal{A}_{TS}(\tau_{exec})$$Specifically, the element $e_t$ at each time step is formalized as a quadruple $\langle p_t, o_t, a_t, f_t \rangle$: \textbf{a) Page Context $p_t$:} The functional semantic category of the current interface (e.g., \textit{Product Details Page}), used to contextualize the agent's state. \textbf{b) Observation $o_t$:} A fine-grained textual description of the visual elements on the current interface (e.g., \textit{A list layout containing titles and prices}). \textbf{c) Operation $a_t$:} A high-level semantic action abstracted from underlying low-level events (e.g., \textit{Click [Submit] button}). \textbf{d) System Feedback $f_t$:} The observable system response triggered by the operation (e.g., \textit{Page Transition} or \textit{Toast Notification}), serving to verify the execution outcome.Furthermore, $\mathcal{A}_{TS}$ is responsible for diagnosing Abnormal Termination behaviors based on state evolution. It identifies Early Termination (exiting before the task logic closes) and Delayed Termination (continuing operations after goal achievement), incorporating these instances directly into the ETR metric.

\textbf{Outcome Verifier} $\mathcal{A}_{OV}$ is tasked with quantifying the extent of user requirement satisfaction. Accepting the complete semantic trajectory sequence $\tau_{exec}^\mathcal{S}$ and the Ground Truth Intent $\mathcal{U}_{gt}$ as inputs, it retrieves evidence of fulfillment across the full temporal scope to generate a Boolean vector, $V_{out}$, indicating the completion status of each atomic requirement:
$$V_{out} = [v_1, v_2, ..., v_n] = \mathcal{A}_{OV}(\tau_{exec}^\mathcal{S}, \mathcal{U}_{gt})$$
Specifically, the $i$-th component of the vector, $v_i$, corresponds to the $i$-th atomic requirement, $r_i$, within $\mathcal{U}_{gt}$. This relationship is governed by the logic $v_i = \mathbb{I}(\exists e_t \in \tau_{exec}^\mathcal{S} : e_t \models r_i)$, where $\mathbb{I}(\cdot)$ denotes the indicator function, taking the value 1 if the condition holds and 0 otherwise. This signifies that a requirement is deemed fulfilled provided that any constituent time step, $e_t$, yields sufficient evidentiary support for its satisfaction.Based on the configuration of $V_{out}$, the task is classified as a Success if all requirements are met, a Failure if no requirements are met, or a Partial Success otherwise.

\textbf{Process Inspector} $\mathcal{A}_{PI}$ evaluates the procedural conformity and efficiency of the execution process. Accepting the complete semantic trajectory sequence $\tau_{exec}^\mathcal{S}$ and the reference standard trajectory $\tau_{ref}$ as inputs, it employs semantic alignment analysis to yield the evaluation result tuple:
$$\langle V_{step}, \Omega^\mathcal{A}_{red} \rangle = \mathcal{A}_{PI}(\tau_{exec}^\mathcal{S}, \tau_{ref})$$
The tuple comprises two key components: \textbf{a) Key Step Hit Status $V_{step}$:} Analogous to the definition of $V_{out}$, the vector $V_{step} = [v_1, v_2, ..., v_m]$ indexes the execution status of the $m$ key steps within the reference trajectory $\tau_{ref}$. Its governing logic is $v_j = \mathbb{I}(k_j \in \text{LCS}_{\cong}(\tau_{exec}^\mathcal{S}, \tau_{ref}))$. This signifies that a step is deemed a hit provided the key action $k_j$ establishes a semantic match with an operation in the observed trajectory while strictly preserving relative temporal ordering. \textbf{b) Redundant Operation Set $\Omega_{red}$:} The set $\Omega_{red} \subset \{1, 2, ..., T\}$ encompasses the indices, $t$, of all time steps classified as invalid operations. By juxtaposing the topological structure of the actual trajectory against the reference, $\mathcal{A}_{PI}$ identifies operations attributable to mis-click corrections, invalid backtracking, or infinite loops, subsequently appending them to $\Omega_{red}$.

\textbf{Interaction Auditor} $\mathcal{A}_{IA}$ specializes in scrutinizing the quality of dialogue within human-agent collaboration, aiming to distinguish between effective communication and ineffective interference. Accepting the Interaction History $H_{int}$ and the Ground Truth Intent $\mathcal{U}_{gt}$ as inputs, it yields the evaluation result tuple:
$$\langle \Omega^\mathcal{I}_{vio}, S_{gain} \rangle = \mathcal{A}_{IA}(H_{int}, \mathcal{U}_{gt})$$
This tuple consists of two primary components:Violation Interaction Set $\Omega^\mathcal{I}_{vio}$: The set $\Omega^\mathcal{I}_{vio} \subset \{1, ..., K\}$ aggregates the indices, $k$, of all invalid inquiry turns. $\mathcal{A}_{IA}$ sequentially evaluates each inquiry, $q_k$, against four violation criteria, appending non-compliant turns to $\Omega^\mathcal{I}_{vio}$: \textbf{a) Repetitive:} Inquiries soliciting parameters already explicitly established within the historical dialogue context or encoded in $\mathcal{S}(\mathcal{I}_{obs})$. \textbf{b) Out-of-Scope:} Inquiries soliciting non-essential information that lies beyond the scope of $\mathcal{U}_{gt}$ and should be determined by the agent's autonomous decision-making. \textbf{c) Context-Missing:} Demanding a user decision without presenting candidate options (e.g., \textit{asking Which one would you like to select? absent of any provided context}). \textbf{d) Trivial Execution:} Soliciting confirmation for low-level UI operations that fall under the agent's autonomous planning responsibilities (e.g., \textit{Should I click the confirm button?}). Information Gain Score ($S_{gain}$): The scalar $S_{gain} \in [0, 1]$ quantifies the cumulative proportion of the initial Cognitive Gap eliminated by the agent throughout the interaction process. It is calculated as $S_{gain} = {|\mathcal{G}_{fill}|}/{|\mathcal{G}|}$, where $\mathcal{G}_{fill}$ denotes the subset of parameters successfully resolved by the agent through valid inquiries within $H_{int}$.

\subsection{Metrics Definition}
Predicated on the quantitative determinations yielded by the aforementioned multi-agent pipeline, MUSE establishes a comprehensive evaluation metric framework spanning three distinct dimensions: Effectiveness, Process Quality, and Interaction Intelligence. Let the evaluation dataset comprise $N$ tasks; for the $i$-th task, we define the relevant parameters as follows:

\subsubsection{Outcome Effectiveness}
This dimension directly assesses the agent's capability to fulfill user intents, derived from the Boolean vector, $V_{out}^{(k)}$, generated by the Outcome Verifier.

\noindent \textbf{Requirement Coverage Rate (RCR):} RCR offers a finer-grained perspective on efficacy compared to binary success rates. It quantifies the average extent to which the agent satisfies the atomic requirements within $\mathcal{U}_{gt}$(encompassing anchors, explicit constraints, and implicit preferences):$$RCR = \frac{1}{N} \sum\nolimits_{k=1}^{N} ({\| V_{out}^{(k)} \|_1}/{| V_{out}^{(k)} |})$$ Here, $|V_{out}^{(k)}|$denotes the total number of atomic requirements associated with the$k$-th task.

\noindent \textbf{Task Success Rate (TSR):} TSR serves as a stringent metric for assessing the holistic completion of a task. A task $k$ is classified as a Success if and only if all its constituent atomic requirements are satisfied (i.e.,$\forall v \in V_{out}^{(k)}, v=1$): $$TSR = \frac{1}{N} \sum\nolimits_{k=1}^{N} \mathbb{I}\big( \min(V_{out}^{(k)}) = 1 \big)$$

\subsubsection{Execution Quality}
This dimension evaluates the procedural conformity, efficiency, and stability of the agent's execution, synthesizing the outputs derived from both the Process Inspector and the Trajectory Serializer.

\noindent\textbf{Key Step Hit Rate (SHR):} SHR quantifies the degree of the agent's adherence to the key path established within the reference trajectory. It is computed based on the hit vector, $V_{step}^{(k)}$, yielded by the LCS-based verification process: $$SHR = \frac{1}{N} \sum\nolimits_{k=1}^{N} ({\| V_{step}^{(k)} \|_1}/{| V_{step}^{(k)} |})$$

\noindent \textbf{Action Redundancy Rate (ARR):} ARR quantifies the proportion of invalid operations exhibited during the agent's execution. Given that the $k$-th task comprises a total of $T^{(k)}$ actual steps:
    $$ARR = \frac{1}{N} \sum\nolimits_{k=1}^{N} \left( {|{{\Omega^\mathcal{A}_{red}}}^{(k)}|} \big/ {T^{(k)}} \right)$$
\noindent \textbf{Error Termination Rate (ETR):} ETR reflects the agent's capacity to govern its own operational boundaries. It is stratified into two distinct categories: Early Termination (ETR-E) and Delayed Termination (ETR-D):
    $$ETR\text{-}X = \frac{1}{N} \sum\nolimits_{k=1}^{N} \mathbb{I}(\text{Term}^{(k)} = \text{Type}_{X})$$
  Here, $X \in \{E, D\}$ specifies the target error category, and$\text{Term}^{(k)}$ denotes the termination type signal associated with the$k$-th task.

\subsubsection{Interaction Quality}
Tailored specifically for human-agent collaborative scenarios, this dimension evaluates the compliance of the dialogue and the efficiency of information acquisition, utilizing the outputs derived from the Interaction Auditor.

\noindent \textbf{Dialogue Compliance Rate (DCR):} DCR assesses the agent's capacity to adhere to established interaction norms. Given that the $k$-th task comprises $K^{(k)}$ interaction turns:
$$DCR = \frac{1}{N} \sum\nolimits_{k=1}^{N} \left( 1 - {|{\Omega^\mathcal{I}_{vio}}^{(k)}|} \big/ {K^{(k)}} \right)$$

\noindent \textbf{Information Gain Rate (IGR):} IGR quantifies the efficiency with which the agent eliminates the Cognitive Gap. It is computed as the mean of the Information Gain Scores ($S_{gain}$) accrued across all interaction turns: $$IGR = \frac{1}{N} \sum\nolimits_{k=1}^{N} S_{gain}^{(k)}$$

\section{Experiment}
In this section, we conduct a comprehensive series of experiments to assess the effectiveness of the AmbiBench dataset and the reliability of the MUSE evaluation framework across multiple dimensions. Our experiments are designed to address the following 3 core Research Questions:
\begin{itemize}
    \item \textbf{RQ1 (Impact of Instruction Clarity):} How does instruction clarity impact the performance boundaries of mobile GUI agents?

    \item \textbf{RQ2 (Execution Quality Diagnosis):} What behavioral patterns and execution defects do fine-grained process metrics reveal?

    \item \textbf{RQ3 (Interaction Efficacy \& Quality):} How effective and high-quality are interactive mechanisms in bridging the cognitive gap?
\end{itemize}

\subsection{Experimental Setup}
\subsubsection{Subject Agents \& Baselines}
To comprehensively assess the capability boundaries across diverse agent paradigms, we selected representative approaches characterizing both Agentic Workflow and Agent-as-a-Model methodologies. The specific agent configurations included in our evaluation are detailed in Table \ref{tab:2}.

\begin{table}[htbp]
\centering
\caption{Overview of subject agents and their average performance across Outcome, Execution, and Interaction dimensions on AmbiBench.}
\label{tab:2}

\resizebox{\columnwidth}{!}{%
\begin{tabular}{lllcccccccccc}
\toprule
\multirow{2}{*}{\textbf{Category}} & \multirow{2}{*}{\textbf{No.}} & \multirow{2}{*}{\textbf{Agentic Framework}} & \multicolumn{2}{c}{\textbf{Model}} & \multirow{2}{*}{\shortstack{\textbf{Support} \\ \textbf{Interaction}}} & \multicolumn{7}{c}{\textbf{Average}} \\
\cmidrule(lr){4-5} \cmidrule(lr){7-13}
 & & & \textbf{Planner} & \textbf{Executor \& Other} & & \textbf{RCR} & \textbf{TSR} & \textbf{SHR} & \textbf{ARR} & \textbf{ETR} & \textbf{DCR} & \textbf{IGR} \\
\midrule
\multirow{3}{*}{\textbf{End-to-End}} & \#1
 & \textbf{Generic Frm.} & \multicolumn{2}{c}{\textbf{UI-Tars-7B} \cite{qin2025uitarspioneeringautomatedgui}} & $\checkmark$ 
 & 24.6 & 21.7 & 50.1 & 4.3 & 72.1 & 87.2 & 12.0 \\
 & \#2 & \textbf{Official Frm.} &\multicolumn{2}{c}{\textbf{AutoGLM-9B} \cite{liu2024autoglmautonomousfoundationagents}}& $\times$ & 48.3 & 35.4 & 59.0 & 28.0 & 54.2 & - & - \\
 & \#3 & \textbf{Generic Frm.} & \multicolumn{2}{c}{\textbf{Qwen-3-VL-8B} \cite{bai2025qwen3vltechnicalreport}}& $\checkmark$ 
 & 17.6 & 10.0 & 36.1 & 53.5 & 90.0 & 88.9 & 2.4 \\
\midrule
\multirow{4}{*}{\textbf{Frameworks}} & \#4
 & \textbf{AppAgent} \cite{appagent} & \multicolumn{2}{c}{\textbf{GPT-4o}} & $\times$ 
 & 27.9 & 10.8 & 42.3 & 36.9 & 72.5 & - & - \\
 & \#5 & \textbf{MobileAgentV2} \cite{mobile-agent-v2} & \textbf{GPT-4o} & \textbf{GPT-4o} & $\times$ 
 & 33.8 & 20.4 & 52.5 & 1.3 & 80.4 & - & - \\
 & \#6 & \textbf{Fairy} \cite{sun2025robustobservableevolvableagentic} & \textbf{GPT-4o} & \textbf{Gemini-3-Flash} & $\checkmark$ 
 & 48.7 & 40.4 & 56.7 & 2.0 & 73.3 & 73.7 & 17.7 \\
 & \#7 & \textbf{Fairy} \cite{sun2025robustobservableevolvableagentic} & \textbf{GPT-4o} & \textbf{UI-Tars-7B} \cite{qin2025uitarspioneeringautomatedgui} & $\checkmark$ 
 & 31.6 & 29.2 & 52.5 & 7.4 & 64.2 & 68.6 & 6.7 \\
\bottomrule
\end{tabular}%
}
\end{table}

\textbf{Generic Agent Framework:} Foundation models within the Agent-as-a-Model paradigm (e.g., GPT-4o, UI-TARS) function essentially as stateless Reasoning Cores, lacking a native runtime environment to maintain the long-horizon context of GUI tasks or manage low-level I/O. Consequently, we constructed a generic agent framework derived from M3A, tasked with assuming control over state management and action execution. This framework serves as the standard Reference Implementation for integration with the MUSE evaluation protocol. To accommodate the interactive nature of AmbiBench, we augmented the original M3A action space with a user interaction action, $a_{ask}$, thereby endowing these models with the capability for active clarification. It is important to note that, due to constraints in instruction-following capabilities, AutoGLM-7B was evaluated exclusively using its official framework.

\textbf{Non-Interactive Agent Frameworks:} Agents belonging to the Agentic Framework paradigm (e.g., AppAgent) inherently encapsulate complete mechanisms for context maintenance and workflow orchestration. To guarantee the Ecological Validity of our evaluation, we adhered to a Non-Interference principle, preserving their original architectural designs. Consequently, if an agent lacks native support for interaction, it is precluded from initiating clarification when confronted with Incomplete or Ambiguous tasks, necessitating reliance on probabilistic decision-making.

\subsubsection{Environment}
\textbf{Device Cluster:} The experiments were conducted on an integrated mobile device cluster server equipped with 20 \textbf{Qualcomm Snapdragon 865} processors, supporting the parallel execution of 20 high-fidelity real-device instances. All instances operated on \textbf{Android 13}, maximally replicating real-world rendering latency and system behaviors to ensure the ecological validity of the evaluation.

\textbf{Inference Deployment \& Configuration:} For open-source models, we utilized their official implementations and deployed them using the vLLM inference framework on high-performance servers equipped with NVIDIA RTX 5090 GPUs. Conversely, for proprietary models such as GPT-4o and UI-TARS, we accessed them exclusively via their official APIs. To mitigate evaluation noise stemming from stochasticity, the Temperature parameter for all models was rigorously set to \textbf{0.0}.

\textbf {MUSE Configuration:} The User Agent within the MUSE framework is empowered by \textbf{GPT-5}. It facilitates necessary parameter clarifications exclusively upon receipt of a compliant $a_{ask}$ request, thereby simulating realistic and rational user behavior. To strike a balance between task complexity and execution efficiency, the Maximum Step Limit $T_{max}$ for each trajectory is constrained to 25 steps, to prevent the task from falling into dead loops (delayed termination) caused by policy failure.

\begin{figure}[h]
  \centering
  \includegraphics[width=\linewidth]{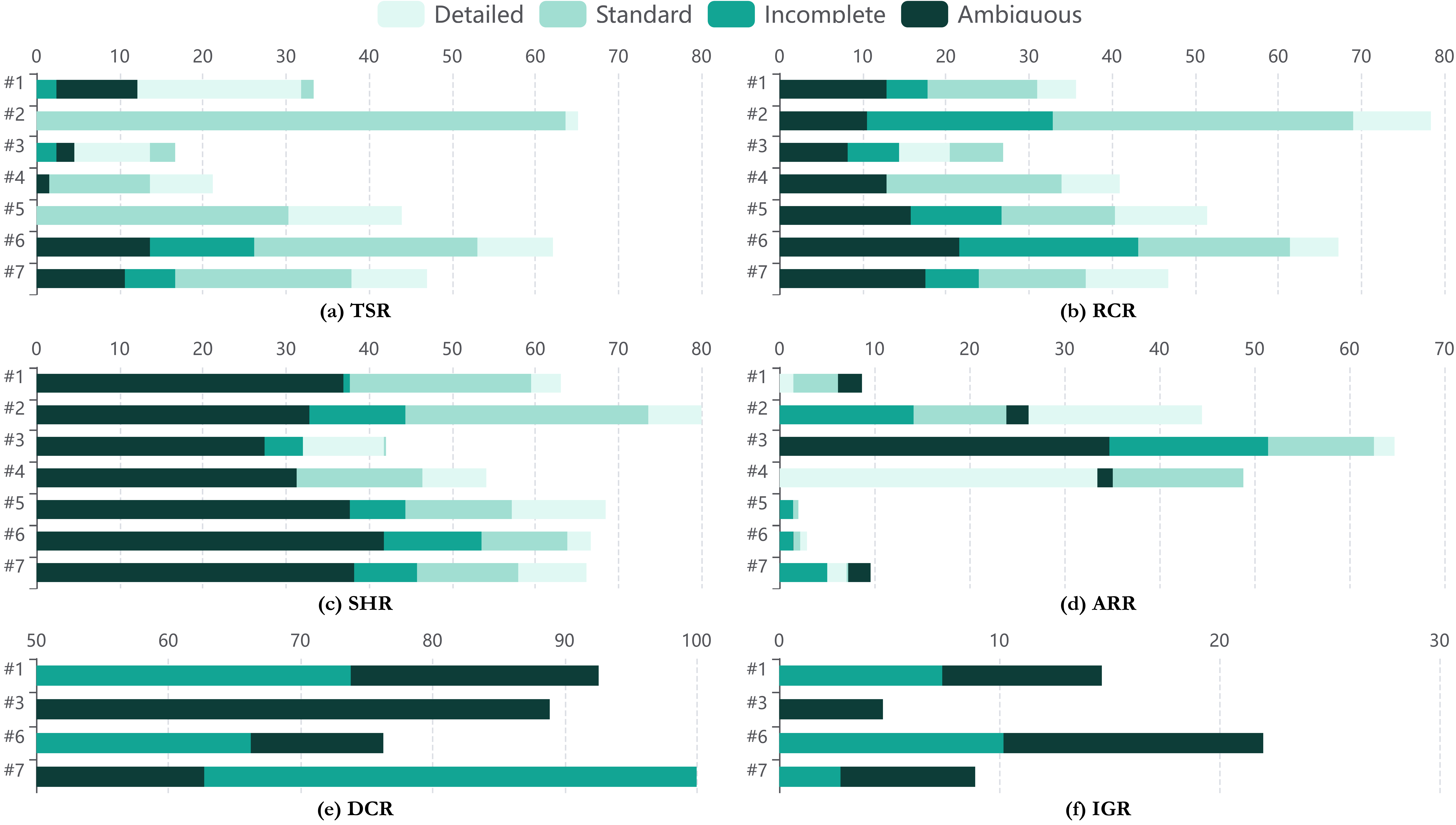}
  \caption{\textbf{Main evaluation results on AmbiBench across three dimensions: Outcome Effectiveness, Execution Quality, and Interaction Quality.}}
  \Description{}
  \label{figure 7}
\end{figure}

\subsection{RQ1: Impact of Instruction Clarity}
To investigate how instruction clarity affects agent performance, we evaluated all baseline agents across the four clarity levels defined in AmbiBench, with the results illustrated in Figure \ref{figure 7} (a)/(b).

Our stratified evaluation exposes a systemic failure of the non-interactive paradigm when confronting real-world ambiguity. In Detailed scenarios, agents such as AutoGLM(\#2) exhibit superior execution proficiency (TSR 65.2\%) by leveraging explicit instructions. However, performance for all non-interactive agents plunges upon reaching the Incomplete and Ambiguous levels. Specifically, in Ambiguous tasks, the TSR for both AutoGLM(\#2) and MobileAgentV2(\#5) drops to zero. This demonstrates that without an interaction mechanism, the unidirectional instruction following paradigm fails to bridge the Cognitive Gap induced by increased information entropy, regardless of the underlying model's reasoning strength.

By leveraging the decoupled design of RCR and TSR in AmbiBench, we identified a deceptive phenomenon where certain agents maintain high RCR and SHR scores in Incomplete and Ambiguous levels despite yielding a TSR of zero. In-depth analysis indicates that this stems not from a genuine comprehension of user intent but from reliance on parametric knowledge to guess partial requirements, resulting in specious outcomes. The experiments further establish interaction capability as a critical watershed for agent performance. Within the interactive environment of AmbiBench, AutoGLM(\#2) and Qwen-3-VL(\#3) exhibit zero willingness to interact and tend to force execution despite information deficits, whereas UI-Tars and Fairy effectively utilize inquiry actions to complete missing information. Results indicate that even agents with weaker baseline execution, such as Fairy(\#6), can surpass SoTA models on ambiguous tasks once interaction is enabled, achieving a score of 26.2\% in Incomplete scenarios. This reversal confirms the necessity of incorporating the interaction dimension in AmbiBench as it successfully shifts the evaluation focus from static execution capability to dynamic alignment capability.

\subsection{RQ2: Execution Quality Diagnosis}

We analyze the experiments for RQ1 using the process quality metrics from AmbiBench, as illustrated in Figure \ref{figure 7} (c)/(d), to diagnose the distinct behavioral characteristics of agents across different architectures when confronting complex tasks.

The incorporation of process data reveals the hidden costs behind apparent success. Although UI-Tars(\#1) and AutoGLM(\#2) both utilize an Agent-as-a-Model architecture, the ARR exposes a fundamental strategic divergence. UI-Tars maintains minimal redundancy of less than 5\% across all clarity levels and demonstrates a precise execution strategy. In contrast, AutoGLM(\#2) exhibits a high redundancy rate of 44.5\% even under Detailed instructions. This comparison highlights the diagnostic value of the process dimension in AmbiBench as it keenly identifies that the high success rate of AutoGLM(\#2) stems partially from trial-and-error exploration rather than rational planning. Such quantification of execution efficiency effectively prevents the overestimation of agent capabilities that occurs when relying solely on TSR.

Our diagnosis of execution processes under varying clarity further pinpoints structural weaknesses in the agents. We observe that even in Detailed scenarios where instructions are most explicit, agents employing the Agentic Workflow architecture lag behind the Agent-as-a-Model AutoGLM(\#2) in terms of SHR. To investigate the causes, we conducted an in-depth diagnosis of the top-performing Fairy(\#6/\#7) series, which trails AutoGLM(\#2) with an SHR of 66.7\% compared to 81.3\%. We trace this counter-intuitive phenomenon to the underlying perception scheme where the UI-Tree-based SoM relies on unreliable XML data in In-the-Wild environments, resulting in missing marks. Without the stratification of clarity and the support of process-level metrics, distinguishing between a perception deviation and a decision failure would remain challenging.

Furthermore, the analysis of abnormal terminations provides insight into the boundaries of agent Meta-cognition. As clarity levels decrease, agents diverge into two distinct failure modes. MobileAgentV2(\#5) and similar models tend towards conservative refusal, characterized by low ARR and immediate termination, which reflects a risk-averse approach to uncertainty. Conversely, AutoGLM(\#2) and Qwen-3-VL(\#3) fall into a pattern of ineffective struggle, exhibiting high ARR and high ETR. This behavior of engaging in extensive random interactions despite the lack of clear guidance reveals that certain agents attempt to hit goal states concealed by instructional ambiguity through blind and inefficient exploration.

\subsection{RQ3: Interaction Efficacy \& Quality}

To explore the value of interaction mechanisms in resolving ambiguity, we analyze the experiments in RQ1 using the interaction quality metrics from AmbiBench, as illustrated in Figure \ref{figure 7} (e)/(f). 

The IGR and DCR reveal a profound contradiction in the behaviors of current agents characterized as being polite but lazy. Data indicates that the majority of interactive agents, such as UI-Tars(\#1), exhibit an exceptionally high DCR, suggesting that they strictly adhere to dialogue norms during the few interactions they trigger. However, a sharp contrast to this high compliance is the extremely low IGR, with Qwen-3-VL(\#3) scoring only 2.4\%, which implies that agents fail to initiate sufficient inquiries. Although current agents possess the Capability for interaction, they severely lack the Awareness to interact as they fail to acutely perceive the ambiguity of current information and active seek clarification. We also observe that while Fairy demonstrates a notable increase in IGR, its DCR declines, indicating a trade-off between Dialogue Compliance and Information Acquisition Effectiveness under current alignment paradigms. Consequently, the core direction for future optimization must focus on enhancing sensitivity to ambiguity and interactive proactivity while simultaneously maintaining high dialogue standards.

Furthermore, to verify the decisive role of interaction mechanisms in ambiguous tasks, we conducted a targeted Mini-Ablation Study. We selected tasks where Fairy(\#6) succeeded in Incomplete scenarios, achieving 100\% in both TSR and RCR, and re-evaluated them with the user interaction function disabled. The results demonstrate that these previously successful tasks immediately failed, with TSR dropping to 0\% and RCR declining to 23.8\%. Due to the singular and significant nature of these findings, we chose not to present them in a separate chart. This phenomenon corroborates the collapse of the non-interactive baselines and establishes a complete chain of evidence that interaction is a necessary condition for agent survival in non-deterministic environments.

\subsection{Reliability Validation}

To validate MUSE’s reliability, three experts independently annotated 100 randomly sampled traces under double-blind conditions, labeling atomic requirement status, key step hit status vectors, and valid parameter fills. High inter-rater agreement (Fleiss’ $\kappa$=0.91) established a solid ground truth. Against this benchmark, MUSE demonstrated strong alignment in $V_{out}$ (Jaccard=0.92) and $V_{step}$ (Jaccard=0.84), while successfully identifying 96\% of valid parameter fills.

\subsection{Threats to Validity}

\subsubsection{External Validity} \textbf{a) Ecological Representativeness:} While AmbiBench covers 240 tasks, its scope remains limited relative to the massive ecosystem of Long-tail Apps. Unique interaction logics in specific vertical domains may not be fully captured, potentially restricting the generalizability of findings to unseen scenarios. \textbf{b) Device Heterogeneity:} Experiments utilized a specific hardware cluster. Given the high fragmentation of the Android ecosystem, variations in ROMs and screen resolutions may impact the generalization of vision-based Mobile GUI Agents across diverse devices.

\subsubsection{Internal Validity} \textbf{a) Non-Determinism in Online Environments:} In-the-wild evaluations inevitably encounter dynamic elements (e.g., A/B testing, real-time ads, network latency). Such environmental noise poses a risk where failures caused by external interference are misattributed to agent incompetence. \textbf{b) Simulator Stability:} Despite strict constraints ($a_{ask}$ triggers), the LLM-driven User Simulator inherently carries risks of hallucination or inconsistency. Occasional unnatural responses could skew Interaction quality metrics (e.g., IGR).

\subsubsection{Construct Validity} \textbf{a) Automated Evaluation Bias}: Although MUSE demonstrates strong correlation with human judgment, MLLM-based auditing may overlook visual nuances or the affective experiences of users in subjective dimensions. \textbf{b) Taxonomy Ambiguity}: The boundaries between clarity levels (particularly Incomplete vs. Ambiguous) can be fluid in language. This potential for subjective interpretation may affect the attribution analysis of behavioral differences.

\section{Conclusion}

To address the oversight of the Cognitive Gap in real-world interactions within existing benchmarks, we propose AmbiBench, the first mobile GUI agent benchmark incorporating a four-level taxonomy of Instruction Clarity, alongside MUSE, a complementary automated auditing framework. Empirical studies reveal the authentic performance boundaries of SoTA agents in scenarios extending beyond one-shot deterministic instructions. Future work will prioritize the construction of high-fidelity local dynamic Apps, an initiative aiming to definitively resolve the Non-Determinism inherent in online evaluation while preserving ecological validity.

\section{Data Availability (This section will not count towards the page limit)}

The AmbiBench benchmark and the MUSE evaluation framework are available in the anonymous GitHub repository: \url{https://anonymous.4open.science/r/MUSE-2973}.

\begin{acks}
\end{acks}

\bibliographystyle{ACM-Reference-Format}
\bibliography{sample-base}

@String{Computing = "Computing" }

@misc{xu2025mobilebenchv2realisticcomprehensivebenchmark,
      title={Mobile-Bench-v2: A More Realistic and Comprehensive Benchmark for VLM-based Mobile Agents}, 
      author={Weikai Xu and Zhizheng Jiang and Yuxuan Liu and Pengzhi Gao and Wei Liu and Jian Luan and Yuanchun Li and Yunxin Liu and Bin Wang and Bo An},
      year={2025},
      eprint={2505.11891},
      archivePrefix={arXiv},
      primaryClass={cs.CL},
      url={https://arxiv.org/abs/2505.11891}, 
}

@misc{song2025colorbenchbenchmarkingmobileagents,
      title={ColorBench: Benchmarking Mobile Agents with Graph-Structured Framework for Complex Long-Horizon Tasks}, 
      author={Yuanyi Song and Heyuan Huang and Qiqiang Lin and Yin Zhao and Xiangmou Qu and Jun Wang and Xingyu Lou and Weiwen Liu and Zhuosheng Zhang and Jun Wang and Yong Yu and Weinan Zhang and Zhaoxiang Wang},
      year={2025},
      eprint={2510.14621},
      archivePrefix={arXiv},
      primaryClass={cs.AI},
      url={https://arxiv.org/abs/2510.14621}, 
}

@misc{mobiagent,
      title={MobiAgent: A Systematic Framework for Customizable Mobile Agents}, 
      author={Cheng Zhang and Erhu Feng and Xi Zhao and Yisheng Zhao and Wangbo Gong and Jiahui Sun and Dong Du and Zhichao Hua and Yubin Xia and Haibo Chen},
      year={2025},
      eprint={2509.00531},
      archivePrefix={arXiv},
      primaryClass={cs.MA},
      url={https://arxiv.org/abs/2509.00531}, 
}

@misc{xing2024understandingweaknesslargelanguage,
      title={Understanding the Weakness of Large Language Model Agents within a Complex Android Environment}, 
      author={Mingzhe Xing and Rongkai Zhang and Hui Xue and Qi Chen and Fan Yang and Zhen Xiao},
      year={2024},
      eprint={2402.06596},
      archivePrefix={arXiv},
      primaryClass={cs.AI},
      url={https://arxiv.org/abs/2402.06596}, 
}

@inproceedings{
rawles2025androidworld,
title={AndroidWorld: A Dynamic Benchmarking Environment for Autonomous Agents},
author={Christopher Rawles and Sarah Clinckemaillie and Yifan Chang and Jonathan Waltz and Gabrielle Lau and Marybeth Fair and Alice Li and William E Bishop and Wei Li and Folawiyo Campbell-Ajala and Daniel Kenji Toyama and Robert James Berry and Divya Tyamagundlu and Timothy P Lillicrap and Oriana Riva},
booktitle={The Thirteenth International Conference on Learning Representations},
year={2025},
url={https://openreview.net/forum?id=il5yUQsrjC}
}

@inproceedings{
wang2025mobileagentbench,
title={MobileAgentBench: An Efficient and User-Friendly Benchmark for Mobile {LLM} Agents},
author={Luyuan Wang and Yongyu Deng and Yiwei Zha and Guodong Mao and Qinmin Wang and Tianchen Min and Wei Chen and Shoufa Chen},
booktitle={The First MARW: Multi-Agent AI in the Real World Workshop at AAAI 2025},
year={2025},
url={https://openreview.net/forum?id=GXugGsJLIP}
}

@inproceedings{
chen2025spabench,
title={{SPA}-{BENCH}: A {COMPREHENSIVE} {BENCHMARK} {FOR} {SMARTPHONE} {AGENT} {EVALUATION}},
author={Jingxuan Chen and Derek Yuen and Bin Xie and Yuhao Yang and Gongwei Chen and Zhihao Wu and Li Yixing and Xurui Zhou and Weiwen Liu and Shuai Wang and Kaiwen Zhou and Rui Shao and Liqiang Nie and Yasheng Wang and Jianye HAO and Jun Wang and Kun Shao},
booktitle={The Thirteenth International Conference on Learning Representations},
year={2025},
url={https://openreview.net/forum?id=OZbFRNhpwr}
}

@inproceedings{
wang2025mobileagente,
title={Mobile-Agent-E: Self-Evolving Mobile Assistant for Complex Tasks},
author={Zhenhailong Wang and Haiyang Xu and Junyang Wang and Xi Zhang and Ming Yan and Ji Zhang and Fei Huang and Heng Ji},
booktitle={Workshop on Scaling Environments for Agents},
year={2025},
url={https://openreview.net/forum?id=GRmrGws6Lf}
}

@misc{kong2025mobileworldbenchmarkingautonomousmobile,
      title={MobileWorld: Benchmarking Autonomous Mobile Agents in Agent-User Interactive and MCP-Augmented Environments}, 
      author={Quyu Kong and Xu Zhang and Zhenyu Yang and Nolan Gao and Chen Liu and Panrong Tong and Chenglin Cai and Hanzhang Zhou and Jianan Zhang and Liangyu Chen and Zhidan Liu and Steven Hoi and Yue Wang},
      year={2025},
      eprint={2512.19432},
      archivePrefix={arXiv},
      primaryClass={cs.CL},
      url={https://arxiv.org/abs/2512.19432}, 
}

@misc{sun2025robustobservableevolvableagentic,
      title={Robust, Observable, and Evolvable Agentic Systems Engineering: A Principled Framework Validated via the Fairy GUI Agent}, 
      author={Jiazheng Sun and Ruimeng Yang and Xu Han and Jiayang Niu and Mingxuan Li and Te Yang and Yongyong Lu and Xin Peng},
      year={2025},
      eprint={2509.20729},
      archivePrefix={arXiv},
      primaryClass={cs.AI},
      url={https://arxiv.org/abs/2509.20729}, 
}

@misc{appagent,
      title={AppAgent: Multimodal Agents as Smartphone Users}, 
      author={Chi Zhang and Zhao Yang and Jiaxuan Liu and Yucheng Han and Xin Chen and Zebiao Huang and Bin Fu and Gang Yu},
      year={2023},
      eprint={2312.13771},
      archivePrefix={arXiv},
      primaryClass={cs.CV},
      url={https://arxiv.org/abs/2312.13771}, 
}

@misc{mobile-agent-v2,
      title={Mobile-Agent-v2: Mobile Device Operation Assistant with Effective Navigation via Multi-Agent Collaboration}, 
      author={Junyang Wang and Haiyang Xu and Haitao Jia and Xi Zhang and Ming Yan and Weizhou Shen and Ji Zhang and Fei Huang and Jitao Sang},
      year={2024},
      eprint={2406.01014},
      archivePrefix={arXiv},
      primaryClass={cs.CL},
      url={https://arxiv.org/abs/2406.01014}, 
}

@misc{AutoDroid,
      title={AutoDroid: LLM-powered Task Automation in Android}, 
      author={Hao Wen and Yuanchun Li and Guohong Liu and Shanhui Zhao and Tao Yu and Toby Jia-Jun Li and Shiqi Jiang and Yunhao Liu and Yaqin Zhang and Yunxin Liu},
      year={2024},
      eprint={2308.15272},
      archivePrefix={arXiv},
      primaryClass={cs.AI},
      url={https://arxiv.org/abs/2308.15272}, 
}

@misc{androidlab,
      title={AndroidLab: Training and Systematic Benchmarking of Android Autonomous Agents}, 
      author={Yifan Xu and Xiao Liu and Xueqiao Sun and Siyi Cheng and Hao Yu and Hanyu Lai and Shudan Zhang and Dan Zhang and Jie Tang and Yuxiao Dong},
      year={2024},
      eprint={2410.24024},
      archivePrefix={arXiv},
      primaryClass={cs.AI},
      url={https://arxiv.org/abs/2410.24024}, 
}

@misc{qin2025uitarspioneeringautomatedgui,
      title={UI-TARS: Pioneering Automated GUI Interaction with Native Agents}, 
      author={Yujia Qin and Yining Ye and Junjie Fang and Haoming Wang and Shihao Liang and Shizuo Tian and Junda Zhang and Jiahao Li and Yunxin Li and Shijue Huang and Wanjun Zhong and Kuanye Li and Jiale Yang and Yu Miao and Woyu Lin and Longxiang Liu and Xu Jiang and Qianli Ma and Jingyu Li and Xiaojun Xiao and Kai Cai and Chuang Li and Yaowei Zheng and Chaolin Jin and Chen Li and Xiao Zhou and Minchao Wang and Haoli Chen and Zhaojian Li and Haihua Yang and Haifeng Liu and Feng Lin and Tao Peng and Xin Liu and Guang Shi},
      year={2025},
      eprint={2501.12326},
      archivePrefix={arXiv},
      primaryClass={cs.AI},
      url={https://arxiv.org/abs/2501.12326}, 
}

@misc{liu2024autoglmautonomousfoundationagents,
      title={AutoGLM: Autonomous Foundation Agents for GUIs}, 
      author={Xiao Liu and Bo Qin and Dongzhu Liang and Guang Dong and Hanyu Lai and Hanchen Zhang and Hanlin Zhao and Iat Long Iong and Jiadai Sun and Jiaqi Wang and Junjie Gao and Junjun Shan and Kangning Liu and Shudan Zhang and Shuntian Yao and Siyi Cheng and Wentao Yao and Wenyi Zhao and Xinghan Liu and Xinyi Liu and Xinying Chen and Xinyue Yang and Yang Yang and Yifan Xu and Yu Yang and Yujia Wang and Yulin Xu and Zehan Qi and Yuxiao Dong and Jie Tang},
      year={2024},
      eprint={2411.00820},
      archivePrefix={arXiv},
      primaryClass={cs.HC},
      url={https://arxiv.org/abs/2411.00820}, 
}

@misc{bai2025qwen3vltechnicalreport,
      title={Qwen3-VL Technical Report}, 
      author={Shuai Bai and Yuxuan Cai and Ruizhe Chen and Keqin Chen and Xionghui Chen and Zesen Cheng and Lianghao Deng and Wei Ding and Chang Gao and Chunjiang Ge and Wenbin Ge and Zhifang Guo and Qidong Huang and Jie Huang and Fei Huang and Binyuan Hui and Shutong Jiang and Zhaohai Li and Mingsheng Li and Mei Li and Kaixin Li and Zicheng Lin and Junyang Lin and Xuejing Liu and Jiawei Liu and Chenglong Liu and Yang Liu and Dayiheng Liu and Shixuan Liu and Dunjie Lu and Ruilin Luo and Chenxu Lv and Rui Men and Lingchen Meng and Xuancheng Ren and Xingzhang Ren and Sibo Song and Yuchong Sun and Jun Tang and Jianhong Tu and Jianqiang Wan and Peng Wang and Pengfei Wang and Qiuyue Wang and Yuxuan Wang and Tianbao Xie and Yiheng Xu and Haiyang Xu and Jin Xu and Zhibo Yang and Mingkun Yang and Jianxin Yang and An Yang and Bowen Yu and Fei Zhang and Hang Zhang and Xi Zhang and Bo Zheng and Humen Zhong and Jingren Zhou and Fan Zhou and Jing Zhou and Yuanzhi Zhu and Ke Zhu},
      year={2025},
      eprint={2511.21631},
      archivePrefix={arXiv},
      primaryClass={cs.CV},
      url={https://arxiv.org/abs/2511.21631}, 
}

@inproceedings{
wang2024mobileagent,
title={Mobile-Agent: Autonomous Multi-Modal Mobile Device Agent with Visual Perception},
author={Junyang Wang and Haiyang Xu and Jiabo Ye and Ming Yan and Weizhou Shen and Ji Zhang and Fei Huang and Jitao Sang},
booktitle={ICLR 2024 Workshop on Large Language Model (LLM) Agents},
year={2024},
url={https://openreview.net/forum?id=jE6pDYCnVF}
}

@misc{ye2025mobileagentv3fundamentalagentsgui,
      title={Mobile-Agent-v3: Fundamental Agents for GUI Automation}, 
      author={Jiabo Ye and Xi Zhang and Haiyang Xu and Haowei Liu and Junyang Wang and Zhaoqing Zhu and Ziwei Zheng and Feiyu Gao and Junjie Cao and Zhengxi Lu and Jitong Liao and Qi Zheng and Fei Huang and Jingren Zhou and Ming Yan},
      year={2025},
      eprint={2508.15144},
      archivePrefix={arXiv},
      primaryClass={cs.AI},
      url={https://arxiv.org/abs/2508.15144}, 
}

@inproceedings{Zhu_2025,
   title={MobA: Multifaceted Memory-Enhanced Adaptive Planning for Efficient Mobile Task Automation},
   url={http://dx.doi.org/10.18653/v1/2025.naacl-demo.43},
   DOI={10.18653/v1/2025.naacl-demo.43},
   booktitle={Proceedings of the 2025 Conference of the Nations of the Americas Chapter of the Association for Computational Linguistics: Human Language Technologies (System Demonstrations)},
   publisher={Association for Computational Linguistics},
   author={Zhu, Zichen and Tang, Hao and Li, Yansi and Liu, Dingye and Xu, Hongshen and Lan, Kunyao and Zhang, Danyang and Jiang, Yixuan and Zhou, Hao and Wang, Chenrun and Zhang, Situo and Sun, Liangtai and Wang, Yixiao and Sun, Yuheng and Chen, Lu and Yu, Kai},
   year={2025},
   pages={535–549} }

@inproceedings{
nong2024mobileflow,
title={MobileFlow: A Multimodal {LLM} For Mobile {GUI} Agent},
author={Songqin Nong and Jiali Zhu and Rui Wu and Jiongchao Jin and Shuo Shan and Xiutian Huang and Wenhao Xu},
booktitle={NeurIPS 2024 Workshop on Open-World Agents},
year={2024},
url={https://openreview.net/forum?id=ylyF4ux9WJ}
}

@misc{hao2025uncertaintyawareguiagentadaptive,
      title={Uncertainty-Aware GUI Agent: Adaptive Perception through Component Recommendation and Human-in-the-Loop Refinement}, 
      author={Chao Hao and Shuai Wang and Kaiwen Zhou},
      year={2025},
      eprint={2508.04025},
      archivePrefix={arXiv},
      primaryClass={cs.AI},
      url={https://arxiv.org/abs/2508.04025}, 
}

@misc{MobileM3,
      title={MobileVLM: A Vision-Language Model for Better Intra- and Inter-UI Understanding}, 
      author={Qinzhuo Wu and Weikai Xu and Wei Liu and Tao Tan and Jianfeng Liu and Ang Li and Jian Luan and Bin Wang and Shuo Shang},
      year={2024},
      eprint={2409.14818},
      archivePrefix={arXiv},
      primaryClass={cs.CL},
      url={https://arxiv.org/abs/2409.14818}, 
}

@misc{li2025autoguiscalingguigrounding,
      title={AutoGUI: Scaling GUI Grounding with Automatic Functionality Annotations from LLMs}, 
      author={Hongxin Li and Jingfan Chen and Jingran Su and Yuntao Chen and Qing Li and Zhaoxiang Zhang},
      year={2025},
      eprint={2502.01977},
      archivePrefix={arXiv},
      primaryClass={cs.CV},
      url={https://arxiv.org/abs/2502.01977}, 
}

@misc{mobile—env,
      title={Mobile-Env: Building Qualified Evaluation Benchmarks for LLM-GUI Interaction}, 
      author={Danyang Zhang and Zhennan Shen and Rui Xie and Situo Zhang and Tianbao Xie and Zihan Zhao and Siyuan Chen and Lu Chen and Hongshen Xu and Ruisheng Cao and Kai Yu},
      year={2024},
      eprint={2305.08144},
      archivePrefix={arXiv},
      primaryClass={cs.AI},
      url={https://arxiv.org/abs/2305.08144}, 
}

@inproceedings{MVISU-Bench,
author = {Huang, Zeyu and Wang, Juyuan and Chen, Longfeng and Xiao, Boyi and Cai, Leng and Zeng, Yawen and Xu, Jin},
title = {MVISU-Bench: Benchmarking Mobile Agents for Real-World Tasks by Multi-App, Vague, Interactive, Single-App and Unethical Instructions},
year = {2025},
isbn = {9798400720352},
publisher = {Association for Computing Machinery},
address = {New York, NY, USA},
url = {https://doi.org/10.1145/3746027.3755695},
doi = {10.1145/3746027.3755695},
booktitle = {Proceedings of the 33rd ACM International Conference on Multimedia},
pages = {8797–8805},
numpages = {9},
location = {Dublin, Ireland},
series = {MM '25}
}

@misc{appagentv2,
      title={AppAgent v2: Advanced Agent for Flexible Mobile Interactions}, 
      author={Yanda Li and Chi Zhang and Wenjia Jiang and Wanqi Yang and Bin Fu and Pei Cheng and Xin Chen and Ling Chen and Yunchao Wei},
      year={2025},
      eprint={2408.11824},
      archivePrefix={arXiv},
      primaryClass={cs.HC},
      url={https://arxiv.org/abs/2408.11824}, 
}

@misc{appagentx,
      title={AppAgentX: Evolving GUI Agents as Proficient Smartphone Users}, 
      author={Wenjia Jiang and Yangyang Zhuang and Chenxi Song and Xu Yang and Joey Tianyi Zhou and Chi Zhang},
      year={2025},
      eprint={2503.02268},
      archivePrefix={arXiv},
      primaryClass={cs.AI},
      url={https://arxiv.org/abs/2503.02268}, 
}

@inproceedings{cocoagent,
    title = "{C}o{C}o-Agent: A Comprehensive Cognitive {MLLM} Agent for Smartphone {GUI} Automation",
    author = "Ma, Xinbei  and
      Zhang, Zhuosheng  and
      Zhao, Hai",
    editor = "Ku, Lun-Wei  and
      Martins, Andre  and
      Srikumar, Vivek",
    booktitle = "Findings of the Association for Computational Linguistics: ACL 2024",
    month = aug,
    year = "2024",
    address = "Bangkok, Thailand",
    publisher = "Association for Computational Linguistics",
    url = "https://aclanthology.org/2024.findings-acl.539/",
    doi = "10.18653/v1/2024.findings-acl.539",
    pages = "9097--9110"
}

@inproceedings{llm-as-a-judge,
title={Judging {LLM}-as-a-Judge with {MT}-Bench and Chatbot Arena},
author={Lianmin Zheng and Wei-Lin Chiang and Ying Sheng and Siyuan Zhuang and Zhanghao Wu and Yonghao Zhuang and Zi Lin and Zhuohan Li and Dacheng Li and Eric Xing and Hao Zhang and Joseph E. Gonzalez and Ion Stoica},
booktitle={Thirty-seventh Conference on Neural Information Processing Systems Datasets and Benchmarks Track},
year={2023},
url={https://openreview.net/forum?id=uccHPGDlao}
}

@misc{a3,
      title={A3: Android Agent Arena for Mobile GUI Agents with Essential-State Procedural Evaluation}, 
      author={Yuxiang Chai and Shunye Tang and Han Xiao and Weifeng Lin and Hanhao Li and Jiayu Zhang and Liang Liu and Pengxiang Zhao and Guangyi Liu and Guozhi Wang and Shuai Ren and Rongduo Han and Haining Zhang and Siyuan Huang and Hongsheng Li},
      year={2026},
      eprint={2501.01149},
      archivePrefix={arXiv},
      primaryClass={cs.AI},
      url={https://arxiv.org/abs/2501.01149}, 
}

@misc{cogAgent,
      title={CogAgent: A Visual Language Model for GUI Agents}, 
      author={Wenyi Hong and Weihan Wang and Qingsong Lv and Jiazheng Xu and Wenmeng Yu and Junhui Ji and Yan Wang and Zihan Wang and Yuxuan Zhang and Juanzi Li and Bin Xu and Yuxiao Dong and Ming Ding and Jie Tang},
      year={2024},
      eprint={2312.08914},
      archivePrefix={arXiv},
      primaryClass={cs.CV},
      url={https://arxiv.org/abs/2312.08914}, 
}

@inproceedings{chi24,
author = {Subramonyam, Hari and Pea, Roy and Pondoc, Christopher and Agrawala, Maneesh and Seifert, Colleen},
title = {Bridging the Gulf of Envisioning: Cognitive Challenges in Prompt Based Interactions with LLMs},
year = {2024},
isbn = {9798400703300},
publisher = {Association for Computing Machinery},
address = {New York, NY, USA},
url = {https://doi.org/10.1145/3613904.3642754},
doi = {10.1145/3613904.3642754},
booktitle = {Proceedings of the 2024 CHI Conference on Human Factors in Computing Systems},
articleno = {1039},
numpages = {19},
location = {Honolulu, HI, USA},
series = {CHI '24}
}

\appendix

\end{document}